\documentclass[12pt,preprint]{aastex}

\begin{document}

\shorttitle{X-ray flare in DQ Tau} \shortauthors{Getman
et al.} \slugcomment{Accepted for publication in ApJ, 01/20/11}

\title{The Young Binary DQ Tau: A Hunt For X-ray Emission From Colliding Magnetospheres}

\author{Konstantin V.\ Getman\altaffilmark{1}, Patrick S.\ Broos\altaffilmark{1}, Demerese M.\ Salter\altaffilmark{2}, Gordon P.\ Garmire\altaffilmark{1}, Michiel R.\ Hogerheijde\altaffilmark{2}}

\altaffiltext{1}{Department of Astronomy \& Astrophysics, 525
Davey Laboratory, Pennsylvania State University, University Park
PA 16802} \altaffiltext{2}{Leiden Observatory, Leiden University, PO Box 9513, 2300 RA Leiden, The Netherlands}

\email{gkosta@astro.psu.edu}

\begin{abstract}
The young high-eccentricity binary DQ~Tau exhibits powerful recurring millimeter-band (mm) flaring attributed to collisions between the two stellar magnetospheres near periastron, when the stars are separated by only $\sim 8$~R$_{\star}$. These magnetospheric interactions are expected to have scales and magnetic field strengths comparable to those of large X-ray flares from single pre-main-sequence (PMS) stars observed in the Chandra Orion Ultradeep Project (COUP). To search for X-rays arising from processes associated with colliding magnetospheres, we performed simultaneous X-ray and mm observations of DQ~Tau near periastron phase. We report here several results. 1) As anticipated, DQ~Tau was caught in a flare state in both mm and X-rays. A single long X-ray flare spanned the entire 16.5 hour $Chandra$ exposure. 2) The inferred morphology, duration, and plasma temperature of the X-ray flare are typical of those of large flares from COUP stars. 3) However, our study provides three lines of evidence that this X-ray flare likely arises from colliding magnetospheres: the chance of capturing a large COUP-like flare within the span of our observation is small; the relative timing of the X-ray and mm flares indicates the Neupert effect and is consistent with a common coronal structure; the size of the emitting coronal structure ($4-5$~R$_{\star}$) inferred from our analysis (which is admittedly model-dependent and should be considered with caution) is comparable to half the binary separation. 4) The peak flare X-ray luminosity is in agreement with an estimate of the power dissipated by magnetic reconnection within the framework of a simple model of interacting magnetospheres.
\end{abstract}


\section{Introduction \label{sec_introduction}}

T-Tauri stars generally show highly elevated levels of X-ray activity arising mostly from violent magnetic reconnection events \citep[e.g.][]{Feigelson99}. This strong X-ray emission has far-reaching implications for the physical processes in the circumstellar environment, the formation of planetary systems, and the evolution of protoplanetary atmospheres \citep[e.g.][]{Glassgold05,Feigelson09}. Recent X-ray surveys of nearby pre-main-sequence (PMS) stellar populations give detailed insights into T-Tauri magnetic flaring; these include the Chandra Orion Ultradeep Project \citep[COUP;][]{Getman05} and the XMM-Newton Extended Survey of Taurus \citep[XEST;][]{Gudel07}. Astrophysical studies of the properties of individual flares and statistical studies of many flares, from both the COUP and XEST observations, reveal that most events are similar to solar magnetic flaring, but with X-ray luminosities up to $10^3-10^5$ times higher than seen in the Sun and plasma temperatures up to $50$ times higher \citep[e.g.][]{Favata05, Wolk05, Flaccomio05, Stassun06, Maggio07, Caramazza07, Franciosini07}. Accretion shocks may contribute only a small fraction to the total X-ray emission from T-Tauri stars, in the form of soft X-ray excess emission \citep[e.g.][]{Telleschi07b, Gudel07b}.

COUP, the deepest and longest (13.2 continuous days) X-ray observation of a young stellar cluster, provided a unique opportunity to study relatively rare (typically $1$ flare per week per star) big X-ray flares from T-Tauri stars \citep{Favata05}. More recently, a detailed and systematic study of $>200$ big flares from $>150$ bright PMS stars detected in the COUP observation by \citet{Getman08a,Getman08b} [hereafter G08a and G08b] shows that they are the most powerful, longest, and hottest stellar flares corresponding to the largest known stellar X-ray coronal structures, reaching up to several stellar radii in both disk-bearing (Class~II) and diskless (Class~III) systems.  The associated large-scale magnetic fields (assuming a dipolar geometry) have an equipartition strength $B = 0.05-0.3$~kG in the outer loop region, consistent with optical Zeeman measurements of surface field strengths around $2-3$~kG in magnetically active T-Tauri stars \citep[e.g.][]{Johns-Krull07,Donati08}. G08ab also provide observational evidence for magnetospheric truncation by a disk in a Class~II system, and for the ability of X-ray loops to withstand centrifugal forces in rapidly rotating Class~III systems. G08ab propose the COUP sample of flares as possible enhanced analogues of very rare Solar Long Decay events (LDEs) associated with X-ray arches and streamers.

While observations of PMS stars at millimeter wavelengths are typically used to study steady thermal emission from dust in their protoplanetary disks, transient gyrosynchrotron and synchrotron continua from flares with spectral peaks in the GHz-THz range can also be seen \citep{Priest02, Kaufmann86}.  Long-term radio variability in older Class~III PMS stars has been known for some time \citep{Garay87} and generally does not show correlations with X-ray variability \citep[e.g.][]{Forbrich07}.  Short-term (hours- or day-long) radio outbursts are occasionally seen from PMS systems, for example: a remarkably powerful millimeter flare from a heavily absorbed Class~III system GMR-A \citep{Bower03};  recurring flares from the Class~III binary V773 Tau A \citep{Massi08} and the Class~II binary system DQ~Tau \citep{Salter08,Salter10};  IRS~5b and IRS~7A Class~I protostars in the Corona Australis cloud \citep{Choi09}; and  a poorly characterized system in Orion's embedded BN/KL star forming region \citep{Forbrich08}. Prior to the current study, only one of these cases, GMR-A, was simultaneously observed with a modern X-ray telescope, by a coincidence with the {\it Chandra}-COUP observation. The GMR-A mm flare was associated, though not exactly simultaneous, with several days of complex big X-ray flares \citep{Bower03, Furuya03, Favata05}. The GMR-A  star is believed to be a single star with a strong coronal magnetic field, while V773 Tau A and DQ Tau are close binary (or triple) systems with component separations at periastron of $30$~R$_{\star}$ and $8$~R$_{\star}$, respectively.  For these multiple systems the magnetic reconnection events have been attributed to interacting magnetospheres. 

The DQ~Tau binary is especially useful in regards to possible X-ray emission. This is a non-eclipsing, double-lined spectroscopic binary, comprised of two relatively equal-mass (equal-radius) PMS stars of $M \sim 0.65$~M$_{\odot}$ ($R \sim 1.6$~R$_{\odot}$) with spectral types in the range of K7 to M1, a rotational period of $P \sim 3$~days for both stars, and a robust orbital period of 15.804 days \citep{Mathieu97,Basri97}. Its highly eccentric orbit ($e=0.556$) exhibits a periastron separation of only $\sim 8$~R$_{\star}$ ($\sim 13$~R$_{\odot}$). The spectral energy distribution (SED) of DQ~Tau is fairly typical of a Class~II system, fit by a large circumbinary disk of about $0.002 - 0.02$~M$_{\odot}$\citep{Mathieu97}. For more than 65\% of periastron encounters, the system experiences optical brightenings as a result of variable and irregular accretion \citep{Mathieu97,Basri97}. The periastron separation is expected to induce magnetospheric interactions at scales and magnetic field strengths comparable to those inferred for the COUP sample of big flares (G08b). We thus performed simultaneous X-ray and mm observations of the orbital segment of DQ~Tau around the peaks of the previously detected mm flares \citep{Salter08} to search for X-ray emission arising from processes associated with colliding magnetospheres. 

We report here a $Chandra$ detection of a long X-ray flare accompanied by mm activity close to the periastron passage of DQ~Tau. The mm observations are discussed in more detail in \citet{Salter10}. The $Chandra$ data are described in \S \ref{sec_chandra_reduction}; and the treatment of mild photon pile-up in the observation is given in \S \ref{sec_pileup_analysis}. Archived {\it XMM-Newton} X-ray observations of DQ~Tau at an orbital phase away from periastron are presented in \S \ref{sec_xmm_reduction}.  Time-integrated $Chandra$ and $XMM-Newton$ spectra are compared in \S \ref{sec_chandravsxmm}. The $Chandra$ flare analyses and the derived flare loop length and loop thickness (within the framework of a single-loop model) are presented in \S \ref{sec_trs} and \S \ref{sec_single_loop_modeling}. A comparison of the $Chandra$ flare with the coincident mm flare is provided in \S \ref{sec_chandra_iram}. A comparison with the COUP sample of big flares is given in \S \ref{sec_coup_star_comparison}. We end in \S \ref{sec_discussion_conclusions} with a discussion of the applicability of the single-loop approach (including information from Appendix ~\ref{sec_solar_flares} on an X-ray analysis of the multi-loop solar X-class flares), our observational findings, and their implications for the origin of the X-ray emission, energetics, and loop geometry.

\section{X-ray Observations and Data Extraction \label{sec_xray_reduction}}

\subsection{$Chandra$ Data \label{sec_chandra_reduction}}

The observation of DQ~Tau was obtained on January $11-12$, 2010 with the ACIS camera \citep{Garmire03} on-board {\it Chandra} \citep{Weisskopf02} as a combined Guest Observer and Guaranteed Time Observation (ObsId~No.~10992, Co-P.I.s: K. Getman \& G. Garmire). The observation was set up based on the expectation of an X-ray flare close to the orbital phase of $\Phi = 0.98$, which corresponds to the peak of the April~2008 large sub-mm flare \citep{Salter08,Salter10}. The total exposure time of the $Chandra$ observation is $\sim 59$~ks, with no data losses or background flaring due to solar activity. The start and end times of the observation are January~11, 2010 at 14$:$02$:$43~UT (JD~2455208.09) and January~12, 2010 at 06$:$31$:$17~UT (JD~2455208.77), respectively. This time period covers the orbital segment of $\Phi = 0.95-0.99$ assuming the orbital parameters, such as time of periastron passage JD$_{0} = 2449582.54$ and orbital period $P = 15.8043$~days, determined by \citet{Mathieu97}. To mitigate photon pile-up effects during an anticipated X-ray flare, the observation was performed with a $1/8$ sub-array of a single ACIS-I3 chip. The aim point of the observation was $04^{\rm{h}}46^{\rm{m}}54\fs2$, $+16\arcdeg59\arcmin35\farcs3$ (J2000), and the satellite roll angle was $284\fdg1$. 

Data reduction follows procedures similar to those described in detail by \citet{Broos10} and \citet[][Appendix~B]{Townsley03}. Briefly, using the tool {\it acis\_process\_events} from the CIAO~4.2 software package, the latest calibration information (CALDB~4.2.0) on time-dependent gain and a custom bad pixel mask are applied, background event candidates are identified, and the data are corrected for CCD charge transfer inefficiency (CTI). Using the {\it acis\_detect\_afterglow} tool, additional afterglow events not detected with the standard Chandra X-ray Center (CXC) pipeline are flagged. The event list is cleaned by ``grade'' (only ASCA grades 0,2,3,4,6 are accepted), ``status'', ``good-time interval'', and energy filters. The slight point-spread function (PSF) broadening from the CXC software position randomizations is removed.

\begin{figure}
\centering
\includegraphics[angle=0.,width=2.0in]{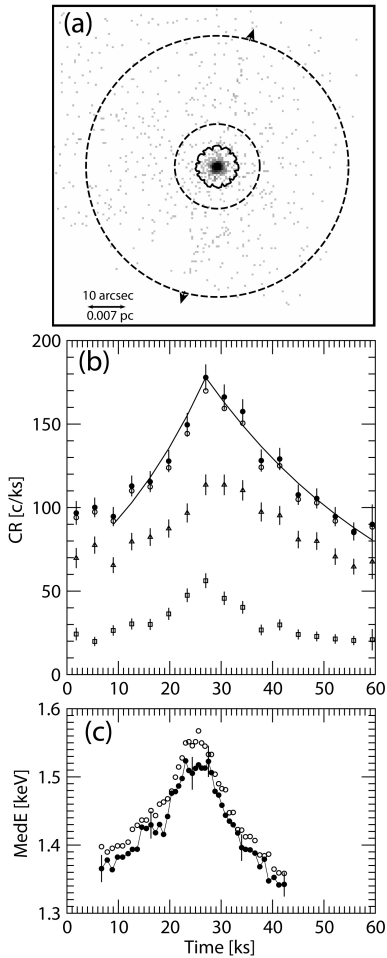}
\caption{\footnotesize The January 2010 59~ks $Chandra$ observation of DQ Tau at the orbital phase $\Phi = 0.95-0.99$ close to periastron passage $\Phi = 0.0, 1.0$. (a) X-ray events detected in a $1.4\arcmin \times 1.4\arcmin$ field around DQ~Tau. Source extraction aperture enclosing $99$\% of the local PSF is marked by the solid ``flower'' contour. Background extraction region (dashed annulus) excludes $> 99$\% of the PSF power. Direction of the faint CCD readout streak emanating from DQ Tau is indicated by the arrowheads. (b) Observed light curves of DQ~Tau binned by 1~hr in the full 0.5--8~keV (open circles), soft $0.5-2$~keV (triangles), and hard $2-8$~keV (squares) energy bands. Light curve in the full band, corrected for mild photon pile-up, (filled circles; \S \ref{sec_pileup_analysis}) with the best-fit exponential rise and decay (solid lines) models ($e$-folding timescales of $\tau_{rise} = 26.3$~ks and $\tau_{decay} = 40.9$~ks, respectively). (c) Evolution of the adaptively smoothed median X-ray event energy, observed (open circles) and corrected for pile-up (filled circles). The smoothing is performed using a sliding boxcar kernel enclosing 1000 counts. For the six independent time segments (segments \#\# 1, 11, 22, 27, 36, 45 from Table~\ref{tbl_trs}) normalized median absolute deviation (MAD) errors on median energy (see Appendix~B in G08a) are indicated by the vertical bars. \label{fig_chandra_image_lc}}
\end{figure}

Using the
\anchor{http://www.astro.psu.edu/xray/acis/acis\_analysis.html}{{\em ACIS Extract}}
(AE) software package
\footnote{
  The {\em ACIS Extract} software package and User's Guide are available at
  \url{http://www.astro.psu.edu/xray/acis/acis\_analysis.html}. 
  } 
\citep{Broos10}, DQ~Tau photons are extracted within a polygonal contour enclosing $\sim 99$\% of the local point spread function (PSF). The background is measured locally in a source-free region (Figure \ref{fig_chandra_image_lc}a). More than 6200 DQ~Tau net counts in the full (0.5--8~keV) band are detected. The AE package was also used to construct source and background spectra, compute redistribution matrix files (RMFs) and auxiliary response files (ARFs), construct light curves and time-energy diagrams, perform a Kolmogorov-Smirnov (K-S) variability test, compute photometric properties, correct spectra and light curves for light pile-up (\S \ref{sec_pileup_analysis}), and  perform automated spectral grouping and fitting of time-resolved data (\S \ref{sec_trs}).

With a fast rise phase (e-folding timescale $\tau_{rise} = 26.3$~ks) and a slow decay phase ($\tau_{decay} = 40.9$~ks), a single long X-ray flare spanned the entire $\sim 16.5$~hour $Chandra$ exposure (Figure~\ref{fig_chandra_image_lc}b). During the observing period the count rate in the full-band (and hard-band) increased by a factor of 2 (and 3), peaking at about $t=27$~ks ($\Phi \sim 0.97$) after the start of the $Chandra$ observation. The evolution of the median energy of X-ray events in the full-band (adaptively smoothed) shows a slower rise and a faster decay than the light curve, reaching a $\sim 5$~ks wide maximum-value plateau at $t \sim 23$~ks, i.e. $4-5$~ks before the count rate peak (Figure ~\ref{fig_chandra_image_lc}b,c). The X-ray count rate and median energy serve as photometric surrogates for emission measure and plasma temperature \citep{Getman10}. There is thus an indication for a time delay between the temperature peak and the emission measure peak (see \S \ref{sec_trs} for an explanation of the effect).

\subsection{$Chandra$ Data Pile-up Analysis \label{sec_pileup_analysis}}

Photon pile-up occurs when two or more photons are incident on a single detector region (for $Chandra$-ACIS this is typically a $3 \times 3$ pixel region) during a single CCD frame. Two or more photons are thus detected as a single event. Some of these pile-up events mimic cosmic rays and are rejected by the on-board processing, while others are telemetered as valid events but with spuriously high energies\footnote{Detailed information on a $Chandra$-ACIS pile-up is given, for example in ``The $Chandra$ ABC Guide to Pileup'', \url{http://cxc.harvard.edu/ciao/download/doc/pileup\_abc.pdf}.}. Despite the use of the $1/8$ sub-array ACIS mode with the reduced CCD frame of 0.5~s (6.4~times shorter than the nominal frame), the $Chandra$ data of DQ~Tau suffer minor pile-up. Assuming a single-temperature plasma with $kT \sim 2$~keV and an average X-ray column density of $N_H \sim 2 \times 10^{21}$~cm$^{-2}$ (\S \ref{sec_char_state}), PIMMS\footnote{\url{http://asc.harvard.edu/toolkit/pimms.jsp}.} gives an estimate on the pile-up fraction (the ratio of the number of frames with two or more events to the number of frames with one or more events) of only $3$\% at the peak of the $Chandra$ flare. Nevertheless, below we adopt two independent approaches to correct the data for pile-up: modeling pileup within the context of spectral fitting using the pile-up model of \citet{Davis01}, and reconstructing a pile-up free spectrum using an experimental Monte Carlo approach of Broos et al. (2010, in prep.). Both methods are applied to the forty five 1000-count overlapping time segments defined in \S \ref{sec_trs}.

In the first approach, for each of the 45 time segments, the pile-up model of Davis is applied along with fixed two-temperature soft component and thawed one-temperature hot (flare) component plasma models subject to constant absorption, as defined and discussed in \S \ref{sec_char_state} and \S \ref{sec_trs}.  According to ``The $Chandra$ ABC Guide to Pileup'', the following parameters are left frozen at their default values: the maximum number of photons to pile-up in a single frame $max_{ph} = 5$, the grade correction for single photon detection $g_0 = 1.0$, and the number of independent $3 \times 3$~pixel pile-up islands in the source extraction region $nregions = 1$. The frame time parameter $frtime$ is set to that of the $Chandra$ observation's frame time of $EXPTIME = 0.5$~s divided by the fractional exposure $FRACEXPO=0.995$. The grade migration parameter ($\alpha$), and the fraction of events in the source extraction region to which pile-up is applied ($psffrac$), are varied within the ranges of $\alpha = [0.5-1.0]$ and $psffrac = [0.90 - 0.95]$, respectively.

In the second approach, an experimental Monte Carlo forward-modeling method to reconstruct unpiled DQ~Tau spectra from piled-up ACIS data is applied. The essential features of this pile-up reconstruction method are briefly described here. A thorough description of the method is given by Broos et al. (2010, in prep). A non-physical ``nuisance model'' with many free parameters feeds an input spectrum to the MARX mirror-detector simulator\footnote{\url{http://space.mit.edu/CXC/MARX/}.} which produces photons that have the correct $Chandra$-ACIS PSF. A physical model of the ACIS CCD produces a piled simulated spectrum. The nuisance model is iteratively adjusted until the piled simulated spectrum is similar to the observed spectrum. The simulation is then run one more time, with pile-up disabled (i.e. exactly one photon arrives per frame). The resulting event list is claimed to be similar to what ACIS itself would have produced if pile-up were not present, and thus one could attempt to fit it with a physical model. For each of the 45 time segments, reconstructed unpiled spectra of DQ~Tau are then fit with fixed two-temperature soft and thawed one-temperature hot plasma component models subject to constant absorption, as defined and discussed in \S \ref{sec_char_state} and \S \ref{sec_trs}.

For all 45 time segments, flare plasma temperature and emission measure inferred by the two methods are consistent within 1$\sigma$. Both methods indicate very light pile-up with $\la 4$\% difference between the observed count rate and that expected if pile-up effects were not present (filled versus open circles in Figure \ref{fig_chandra_image_lc}). Further in the flare analysis (\S \ref{sec_flare_analysis}) we adopt the results from the Broos et al. method.

\subsection{{\it XMM-Newton} Data \label{sec_xmm_reduction}}

\begin{figure}
\centering
\includegraphics[angle=0.,width=4.in]{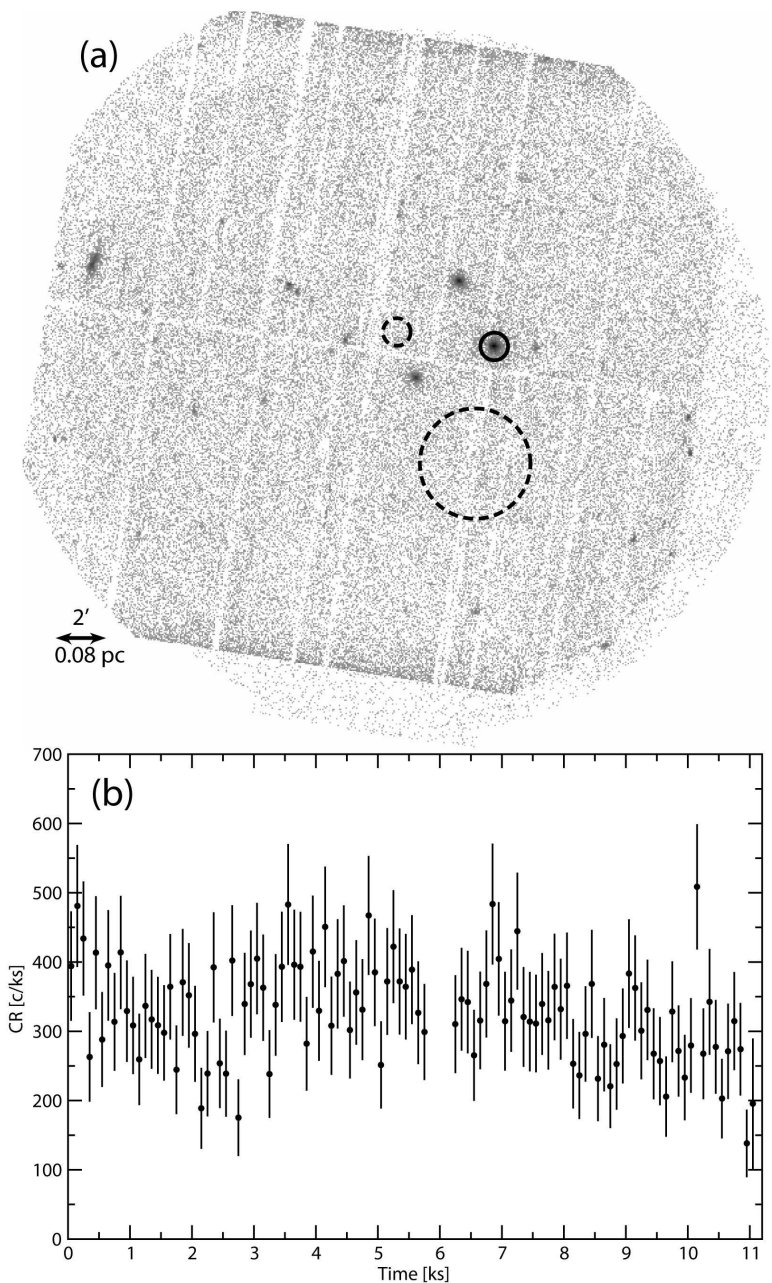}
\caption{The February 2007 12.6~ks {\it XMM-Newton} observation of DQ Tau at the orbital phase $\Phi \sim 0.65-0.7$. (a) Merged EPIC MOS+PN $(0.2-10)$~keV image of the $\sim 15\arcmin \times 15\arcmin$ region around DQ~Tau. The extraction aperture (small solid circle) encloses 90\% of the local PSF power. MOS and PN background extraction regions are marked by the large and small dashed circles, respectively. (b) Background subtracted EPIC-PN light curve of DQ Tau in the $(0.2-10)$~keV energy range with a bin size of 100~s. \label{fig_xmm_image_lc}}
\end{figure}

We use archived, publicly available {\it XMM-Newton} \citep{Jansen01} observations of DQ~Tau. As part of the program aimed at studying X-ray properties of EX Lupi-type outburst (EXOR) stars, the EXOR star DR~Tau was observed by {\it XMM-Newton} on February~13, 2007 (ObsId No.~0406570701, P.I. G. Stringfellow). DQ~Tau is $3.3\arcmin$ north-west of DR~Tau, within the field of view of the {\it XMM-Newton} EPIC \citep{Struder01,Turner01} detectors (Figure \ref{fig_xmm_image_lc}a). Both EPIC-MOS and EPIC-PN cameras were configured in full window mode with medium and thin filters for MOS~1,2 and PN cameras, respectively. The total exposure time for both EPIC-MOS~1,2 detectors is $12.6$~ks, and it is $11$~ks for the EPIC-PN. The observation start and end times, 04$:$20$:$12~UT (JD~2454144.68) and 07$:$50$:$28~UT (JD~2454144.83) respectively, correspond to a time period covering the $\Phi = 0.66-0.67$ orbital segment of DQ~Tau.

Our data reduction follows procedures described in the Common and EPIC related Science Analysis Software (SAS) Threads\footnote{The SAS Threads are available at
  \url{http://xmm.esac.esa.int/sas/current/documentation/threads/}.}. Briefly, the data were reprocessed with the SAS version 9.0.0. The {\it emproc} and {\it epproc} meta-tasks were run to obtain calibrated and concatenated event lists for EPIC-MOS and EPIC-PN detectors, respectively. Short intervals of flaring particle background of $100$~s ($400$~s) long around the observation time of $t = 6$~ks for MOS (PN) detectors were identified and removed. The event lists were further cleaned with the PATTERN, FLAG, and energy filters\footnote{The following filters recommended in \url{http://xmm.vilspa.esa.es/docs/documents/CAL-TN-0018.pdf} are used. For both the EPIC-MOS and EPIC-PN detectors, the energy filter of $(0.2-10)$~keV is applied. For the EPIC-PN detector, excluding border pixels (FLAG == 0), single- and double-pixel events (PATTERN $<$=4) flagged as ``good'' (\#XMMEA\_EP) are retained. For the EPIC-MOS detector single-, double-, triple-, and quadruple-pixel events (PATTERN $<$=12) flagged as ``good'' (\#XMMEA\_EM) are retained.}.

The DQ~Tau photons were extracted within a $33.5\arcsec$ radius circular region ($\sim 90\%$ encircled PSF). For the MOS~1,2 images, background was measured on the same CCD away from the source; for the PN image, a background region was chosen from a source-free region on a neighboring CCD at the same distance to a corresponding readout node (Figure \ref{fig_xmm_image_lc}a). More than 4100 DQ~Tau EPIC net counts in the $(0.2-10)$~keV band were detected. The SAS tasks {\it evselect} and {\it epatplot} suggest that the EPIC image of DQ~Tau is not affected by pile-up.

Following the SAS threads, the SAS tasks {\it evselect}, {\it backscale}, {\it rmfgen}, {\it arfgen}, {\it epiclccorr} and a number of tools from HEASOFT version 6.8 were used to construct source and background spectra, compute RMFs and ARFs, and construct light curves. During the {\it XMM} observation, DQ~Tau showed no significant variations in its X-ray light curve (Figure~\ref{fig_xmm_image_lc}b).

\section{$Chandra$ Flare Analysis \label{sec_flare_analysis}}

\subsection{Time-integrated $Chandra$ spectrum \label{sec_char_state}}

The time-integrated $Chandra$ spectrum is modeled here with a three-temperature APEC plasma emission model \citep{Smith01}. The APEC model is also employed in the time-resolved flare spectroscopy (\S \ref{sec_trs}).  However, for a better modeling of individual spectral features, comparison between the $Chandra$ and {\it XMM-Newton} data in \S \ref{sec_chandravsxmm} is performed using the VAPEC model \citep{Smith01}.

The fit to the time-integrated $Chandra$ spectrum is performed using $\chi^{2}$ statistics. The light $Chandra$ data pile-up (\S \ref{sec_pileup_analysis}) is ignored here. In the fit, elemental abundances are frozen at 0.3 times solar, which has been previously suggested as typical values for young stellar objects \citep{Imanishi01,Feigelson02}. Solar abundances are taken from \citet{Anders89}. X-ray absorption is modeled using the atomic cross sections of \citet{Morrison83}. The fit is performed with the XSPEC spectral fitting package version 12.5.1n \citep{Arnaud96}.

The X-ray spectral fitting of the DQ-Tau data is ambiguous. At least two qualitatively different spectral model families give similarly good values of reduced $\chi^{2}_\nu \sim 0.9$\footnote{Ambiguity in spectral fitting of hundreds of bright X-ray young stellar objects was previously reported by \citet{Getman05}.} (see Figure~\ref{fig_chandra_xmm_spectra}a,b with analogous VAPEC fits). The best fit parameters for the first model family are around $kT_1 = 0.7$~keV, $kT_2 = 1.9$~keV, $kT_3 = 4.3$~keV, $N_H = 1.3 \times 10^{21}$~cm$^{-2}$ (corresponding to a visual extinction of $A_V \sim 0.8$~mag using the gas-to-dust relationship of \citet{Vuong03}), and for the second family are around $kT_1 = 0.3$~keV, $kT_2 = 0.9$~keV, $kT_3 = 3.5$~keV, $N_H = 3.2 \times 10^{21}$~cm$^{-2}$ ($A_V \sim 2$~mag). The second solution gives softer temperatures and higher absorption. The high temperature model component ($kT_3$) is assumed here to be fully associated with the X-ray emission from the flare. A summary of the low temperature model components ($kT_1$ and $kT_2$) for both solutions is given in Table \ref{tbl_ch_models}. Both seem physically reasonable. For example, in the first solution, the temperatures ($kT_1$ and $kT_2$) and the ratio of emission measures ($EM_2/EM_1$) are consistent with those for COUP PMS stars \citep[Table~1 in][]{Getman10}. In the second solution, $kT_1 \sim 0.3$~keV might represent a soft excess emission due to accretion \citep[e.g.][]{Telleschi07b}; DQ~Tau is known to experience pulsed accretion flows near periastron \citep{Basri97}. For both solutions, the absorption-corrected total-band (0.5--8~keV) luminosities of the low temperature component ($kT_1$ and $kT_2$) lie within the locus of XEST Class~II stars on the X-ray luminosity versus mass diagram \citep{Telleschi07a}. Finally, the X-ray column densities from both solutions are consistent with the range of DQ~Tau visual extinction estimates given in the literature, from $A_V = 0.5$~mag \citep{Cohen79} to $A_V = 2.1$~mag \citep{Strom89}. Thus, in the following analysis of the $Chandra$ flare we will consider both solutions, which we will refer to as low temperature plasma component Model~1 and Model~2.

\subsection{X-ray Emission at a Different Epoch and Phase \label{sec_chandravsxmm}}

\begin{figure}
\centering
\includegraphics[angle=0.,width=6.5in]{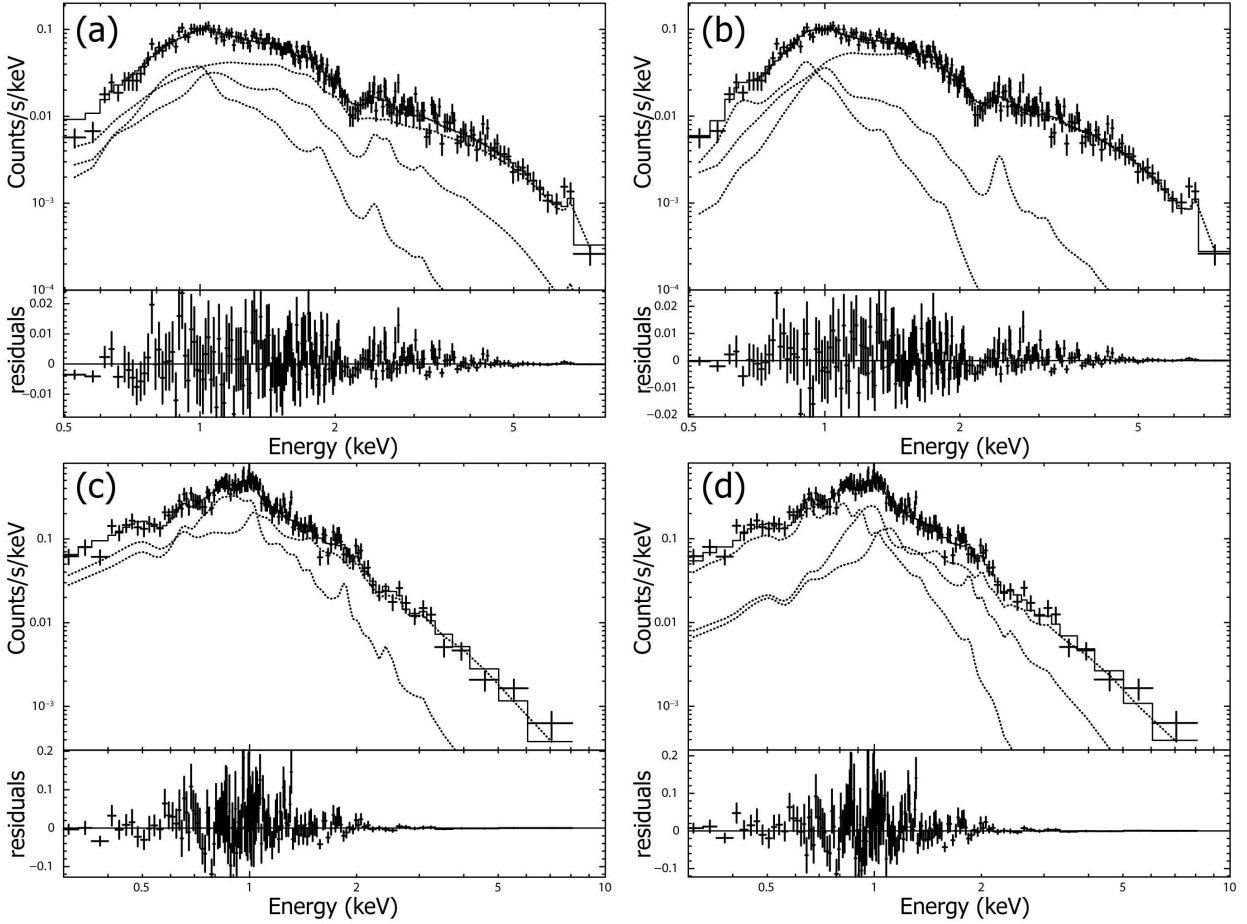}
\caption{Comparison of the time-integrated $Chandra$-ACIS and {\it XMM-Newton}-EPIC spectra, and the ambiguity in the spectral fitting. Panels (a) and (c) show spectral fits to the $Chandra$ and {\it XMM-Newton} data, respectively, with their $N_H$, $kT_1$, $kT_2$, $EM_2/EM_1$ parameters fixed at values of the low temperature plasma component Model~1 (Table~\ref{tbl_ch_models}). $Chandra$ modeling requires an additional time-averaged flare component with $kT_3 = 4.3$~keV. Panels (b) and (d) show the $Chandra$ and {\it XMM-Newton} fits, respectively, with $N_H$, $kT_1$, $kT_2$, $EM_2/EM_1$ fixed at values of the low temperature plasma component Model~2 (Table~\ref{tbl_ch_models}). $Chandra$ modeling requires an additional time-averaged hot temperature (flare) component with $kT_3 = 3.5$~keV. {\it XMM-Newton} modeling requires an additional moderately hot component $kT_3 = 1.9$~keV. \label{fig_chandra_xmm_spectra}}
\end{figure}

Compared to the emission near periastron ($\Phi = 0.95-0.99$) observed in January~2010 by {\it Chandra}, the emission away from periastron ($\Phi = 0.66-0.67$) observed in February~2007 by {\it XMM-Newton} is rather constant (Figure~\ref{fig_xmm_image_lc}b) and, as shown below, softer. The time-integrated $Chandra$-ACIS and {\it XMM-Newton}-EPIC\footnote{This is a combined PN$+$MOS1,2 spectrum created following the SAS thread \url{http://xmm.esac.esa.int/sas/current/documentation/threads/epic\_merging.shtml}.} spectra are compared in Figure \ref{fig_chandra_xmm_spectra}. Both spectra are fit with a two- or three-temperature VAPEC plasma emission model with individual elemental abundances set to values typical for PMS and for extremely active zero-age main-sequence (MS) stars, as specified in \citet{Gudel07}\footnote{With respect to the solar photospheric abundances of \citet{Anders89}, individual elemental abundances adopted in our spectral fits are: C$=$0.45, N$=$0.788, O$=$0.426, Ne$=$0.832, Mg$=$0.263, Al$=$0.5, Si$=$0.309, S$=$0.417, Ar$=$0.55, Ca$=$0.195, Fe$=$0.195, Ni$=$0.195.}. The {\it XMM-Newton} spectrum can be successfully fit by the low temperature plasma component Model~1 with no need for a hotter component, but with a flux 1.5 times higher than that of the low temperature plasma emission in the $Chandra$ spectrum (panels (c) versus (a) in Figure \ref{fig_chandra_xmm_spectra}). Alternatively, the {\it XMM-Newton} spectrum can be fit with the low temperature plasma component Model~2 with a flux 1.7 times lower than that of the low temperature emission in the $Chandra$ spectrum, and an additional moderately hot component ($kT_3 = 1.9$~keV) with a flux 4.5 times lower than that of the $Chandra$ hot component ($kT_3 = 3.5$~keV) (panels (d) versus (b)). Differences in plasma emission between the two epochs, spaced roughly 3 years apart, could be simply attributed to evolution of magnetic active regions on the stellar surface. The 1.5-fold flux difference between the {\it XMM-Newton} spectrum and the Model~1 low temperature component emission in the $Chandra$ spectrum has an alternate, less likely, explanation --- the known suppression of time-integrated X-ray emission in accreting versus non-accreting stars \citep[e.g.][]{Flaccomio03,Preibisch05}; phase-variable mass flows from the circumbinary disk with accretion rates $10-200$ times higher near periastron than those away from periastron are proposed to take place in the DQ~Tau system \citep{Mathieu97,Basri97}.

\subsection{Time-Resolved Spectroscopy \label{sec_trs}}

As in G08a, flare spectral modeling is performed here on multiple adjacent overlapping time segments. Each time segment is specified by moving a rectangle (``boxcar'') kernel of variable width through the $Chandra$ event time series to encompass 1000 X-ray counts. Start time of each segment is specified to be offset from the start time of the previous segment by 0.1 times the duration of the previous segment. Forty five segments are defined. The typical width of the smoothing kernel is 7~ks at the peak of the flare and 10~ks at the base of the flare (Table~\ref{tbl_trs}). Among the 45 overlapping segments, sets of 5-6 independent segments can be identified.  Examining dozens of overlapping time segments along the decay phase of a flare light curve, rather than just a handful of independent intervals, allow higher time resolution which often results in the discovery of a more detailed, often more complex flare behavior (Appendix~A in G08a).

\begin{figure}
\centering
\includegraphics[angle=0.,width=6.5in]{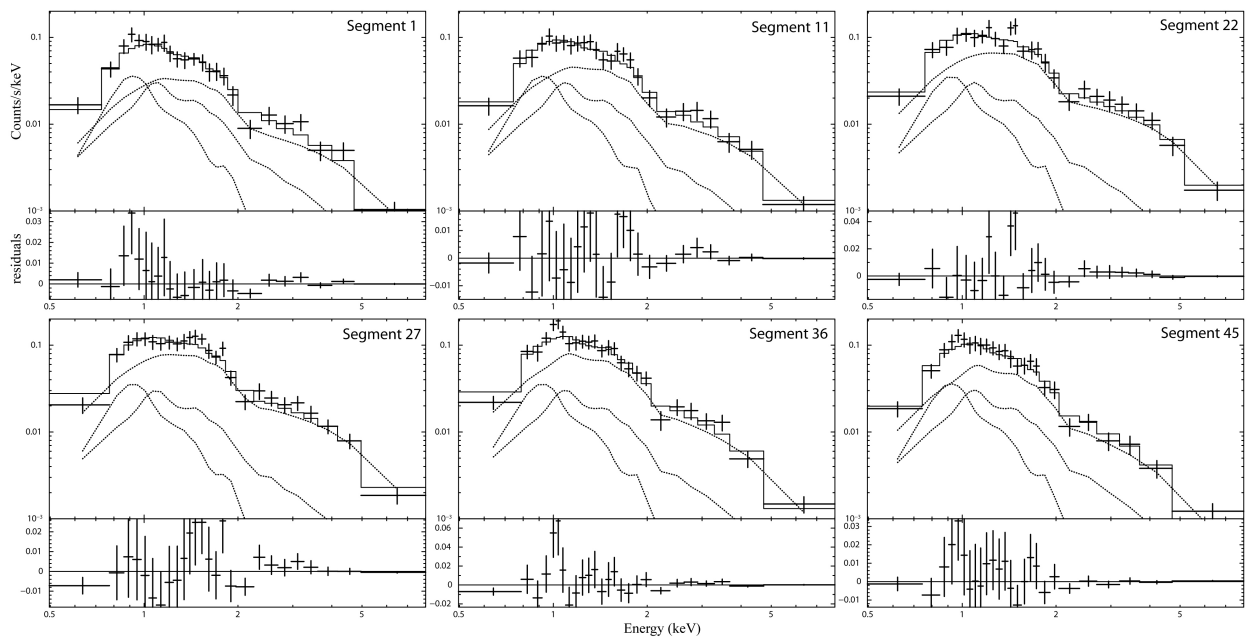}
\caption{Fits of the pile-up corrected $Chandra$ spectra for the six independent time segments (segments \#\# 1, 11, 22, 27, 36, 45 from Table~\ref{tbl_trs}) exemplify our time-resolved spectral modelling. For each of the segments the 2-T soft plasma component model is frozen (two leftmost dotted lines) while the plasma temperature and emission measure of the 1-T hot (flaring) component (rightmost dotted line) are varied to obtain the best-fit. Modeling is shown for the case of the low temperature plasma component Model~1 (Table~\ref{tbl_ch_models}). \label{fig_trs_examples_model1}} 
\end{figure}

Both piled and unpiled spectra (\S \ref{sec_pileup_analysis}) from each segment are fit in XSPEC with thawed one-temperature APEC hot component (flare) and fixed two-temperature APEC cooler component (Table~\ref{tbl_ch_models}) models subject to constant WABS absorption. As in \S \ref{sec_char_state}, elemental abundances are set to 0.3 times solar, and fits are performed using $\chi^{2}$ statistics. Examples of the spectral modeling are shown in Figure~\ref{fig_trs_examples_model1}, and the unpiled spectral results (plasma temperature, emission measure, corrected-for-absorption X-ray luminosities) for both Model~1 and Model~2 scenarios are reported in Table~\ref{tbl_trs}.

\begin{figure}
\centering
\includegraphics[angle=0.,width=6.5in]{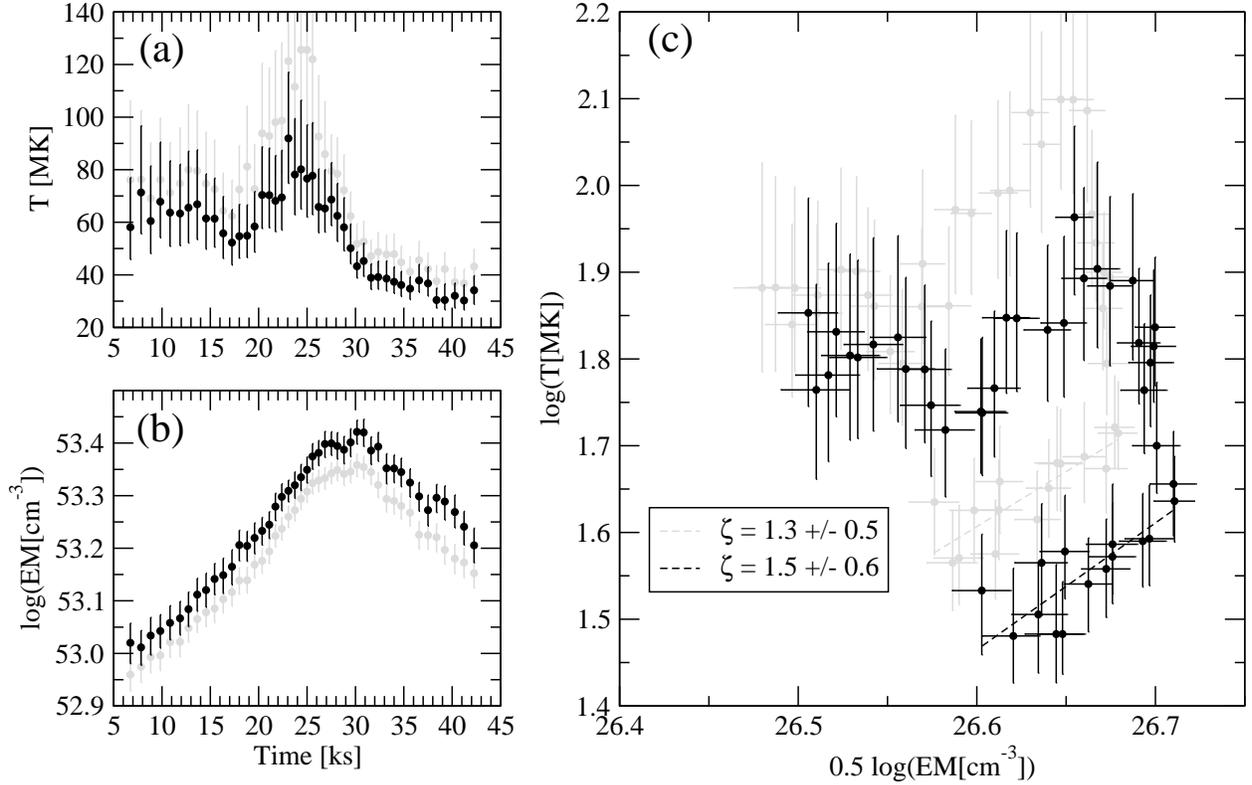}
\caption{Evolution of the inferred plasma temperature (a) and emission measure (b) of the $Chandra$ DQ Tau flare using low temperature plasma component Model~1 (see Table~\ref{tbl_trs}). (c) Evolution in the $\log T$~--~$\log \sqrt{EM}$ plane with the best-fitting flare decay as a dashed line of slope $\zeta$. Black (and gray) points are shown for the cases of the pile-up corrected (and uncorrected) data. Temperature and emission measure points are derived for overlapping time segments (\S \ref{sec_trs}), thus their errors are not independent. \label{fig_comparison_pileup_nonpilep_model1}} 
\end{figure}

\begin{figure}
\centering
\includegraphics[angle=0.,width=6.5in]{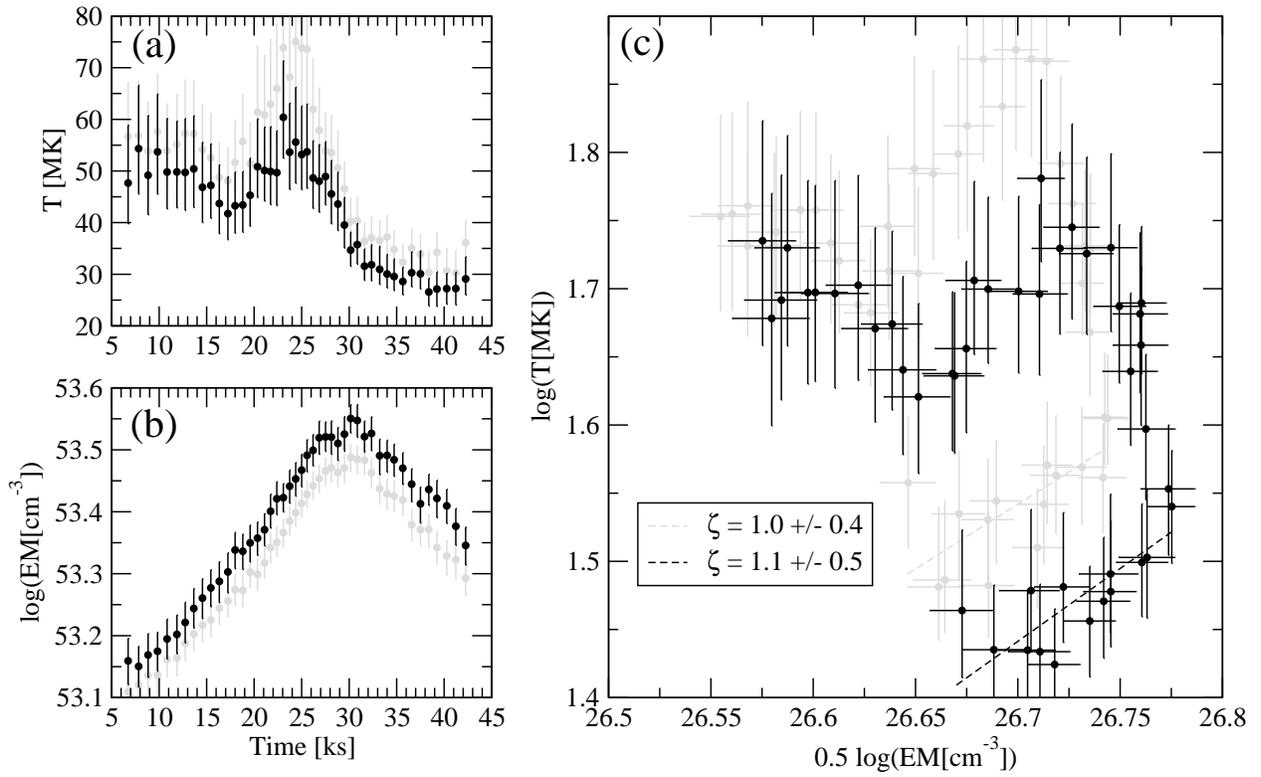}
\caption{Evolution of the inferred $Chandra$ DQ Tau flare properties using low temperature plasma component Model~2. See Figure~\ref{fig_comparison_pileup_nonpilep_model1} for details. \label{fig_comparison_pileup_nonpilep_model2}} 
\end{figure}

For Model~1 and Model~2, respectively, Figures \ref{fig_comparison_pileup_nonpilep_model1} and \ref{fig_comparison_pileup_nonpilep_model2} give the inferred temporal evolution of the plasma temperature and the emission measure, as well as flare evolution in the $\log T - \log \sqrt{EM}$ plane. Without pile-up correction (gray points), flare temperatures would have been overestimated by $20-25$\% and emission measures would have been underestimated by $10$\%. During the first 15-20~ks, the temperature profile shows a plateau with a relatively constant plasma temperature around $50-60$~MK, followed by a drop to $40-50$~MK. This might be an indication of a precursor flare event. Afterwards, the temperature profile is typical of many big flares from the COUP sample of flares: starting with a $5$~ks period where the temperature rapidly rises to $60-90$~MK, then in the next $5$~ks it rapidly drops to $30-40$~MK, and decreases more gradually to $25-30$~MK over the following $>10$~ks.

A $5-8$~ks time delay is seen between the temperature peak and the emission measure peak. Such delays are often observed in solar \citep[e.g.,][and Appendix~\ref{sec_solar_flares} of this work]{Milkey71,Gudel96}, stellar \citep[e.g.,][and references therein]{Reale07}, and flares from PMS stars (e.g., Table~1 in G08a). The delay can be interpreted as an efficient cooling of the plasma by conduction and radiation while the emission measure is still rising due to ongoing evaporation in the loop \citep{Gudel96}. The effect is applicable to both flares from single and multiple loops (see Appendix~\ref{sec_solar_flares}).

The derived slope of the trajectory in the DQ Tau flare's temperature-density diagram $\zeta = 1.1-1.5$ indicates mild or no sustained heating in the framework of a single-loop flaring model (see \S \ref{sec_loop_length} for more details).

\subsection{Single-Loop Modeling \label{sec_single_loop_modeling}}

\subsubsection{Flare Loop Length \label{sec_loop_length}}

In order to derive the size of the coronal structure associated with the $Chandra$ flare in DQ~Tau, we employ the time-dependent hydrodynamic model of \citet{Reale97}[hereafter R97] for a single dominant coronal magnetic loop.  Those authors  establish a formula for estimating a loop's half-length $L$ (throughout the text the terms ``size'' or ``length'' will be used to indicate the loop's half-length) accounting for the possibility of prolonged heating during the decay phase. They find
\begin{equation}
L = \frac{\tau_{decay} \sqrt{T_{pk}}}{3.7 \times 10^{-4} F(\zeta)}
\label{eqn_loopsize}
\end{equation}
where $L$ is the half-length of the loop (cm), $\tau_{decay}$ is the flare decay e-folding timescale (sec), and $T_{pk}$ is the loop apex flare temperature (K) at the time of the maximum emission measure. $F(\zeta)$ is a correction factor for prolonged heating that is a function of the slope $\zeta$ of the trajectory in the temperature-density diagram. In practice, $F(\zeta)$ and $T_{pk}$ must be calibrated for each X-ray observatory; the slope $\zeta$ is usually measured in the $\log T - \log (EM^{1/2})$ plane where $EM$ is the evolving emission measure and $EM^{1/2}$ is used as a proxy for the plasma density. We reproduce the calibration formulas for $F(\zeta)$ and $T_{pk}$ derived for $Chandra$-ACIS by \citet{Favata05}
\begin{equation}
F(\zeta) = \frac{0.63}{\zeta-0.32} + 1.41
\label{eqn_fzeta}
\end{equation}
and
\begin{equation}
T_{pk} = 0.068 \times T_{obs}^{1.2}
\label{eqn_peak_temperature}
\end{equation}
where $T_{pk}$ is a temperature at the loop apex (K), and $T_{obs}$ is an observed ``average'' loop temperature (K) obtained from our {\it Chandra}-ACIS data. The $\zeta \simeq 1.5$ corresponds to a freely decaying loop with no sustained heating, while $\zeta \sim 0.32$ corresponds to a loop with a long sustained heating.

This model is simplistic in a number of ways.  It assumes that the plasma has a uniform density with a unity filling factor confined within a single semicircular loop of uniform cross-section. Furthermore, the model assumes that this geometry remains unaltered during the flare, that energy is efficiently transported along the magnetic field lines of the loop, and that there is continuous  energy balance between the loop heating and the thermal conduction and radiative losses. Despite these limitations, the R97 model has been applied to a variety of solar and stellar flares, including flares from PMS stars in the ONC and Taurus molecular cloud star-forming regions studied by \citet{Favata05, Getman08a, Franciosini07}.  The model has the advantage over earlier and simpler cooling loop models \citep[e.g.][]{Rosner78,Serio91}, which neglect reheating during the decay phase and thereby tend to overestimate loop sizes. Note, however, the limitations of the single-loop approach and its application to the DQ Tau flare discussed in \S \ref{sec_single_vs_multiple}.

The decay timescale inferred for the DQ Tau flare is $\tau_{decay} = 40.9$~ks (\S \ref{sec_xray_reduction}). In the case of the Model~1 (Model~2) the observed ``average'' loop temperature at the time of the peak of the emission measure (time segment \# 27 in Table \ref{tbl_trs}) is $T_{obs} =  68.6_{- 9.9}^{+14.0}$~MK ($T_{obs} =  48.9_{- 5.5}^{+ 6.7}$~MK), and the slope of the trajectory in the temperature-density diagram is $\zeta = 1.5 \pm 0.6$ ($\zeta = 1.1 \pm 0.5$; see Figures \ref{fig_comparison_pileup_nonpilep_model1}-\ref{fig_comparison_pileup_nonpilep_model2}). Applying formulas (\ref{eqn_loopsize}) - (\ref{eqn_peak_temperature}) we find the size of the DQ Tau flaring coronal structure $L = (5.8 \pm 0.7) \times 10^{11}$~cm ($L = 5.2 \pm 0.6$~R$_{\star}$) and $L = (4.4 \pm 0.6) \times 10^{11}$~cm ($L = 4.0 \pm 0.6$~R$_{\star}$) for the Model~1 and Model~2, respectively. This size is comparable to independently inferred heights for the mm emitting regions of $3.7-6.8$~R$_{\star}$ above the stellar surface \citep{Salter10}. The loop size of $L = 4-5$~R$_{\star}$ is also comparable to half the $\gtrsim 8$~R$_{\star}$ component separation near the periastron passage \citep{Mathieu97}.

The R97 model assumes that loop growth is unrestrained by gravity. This is clearly the case for DQ~Tau flare, where the derived half-length of the flaring loop is much shorter than the pressure scale height\footnote{The pressure scale height is $H_p = (k_b T_{obs})/(\mu m_p g)$. The gravitational acceleration near the stellar surface of either DQ~Tau component, $g$, is $0.25 \times g_{\odot} = 6900$~cm~s$^{2}$. $m_p$ is the proton mass and $\mu m_p = 0.6 \times m_p$ is an average mass of a particle in a fully ionized solar plasma.}, $H_p$, both near the stellar surface ($H_p \sim 7-16$~R$_{\star}$) and at a few stellar radii from the surface ($H_p \ga 100$~R$_{\star}$).

\subsubsection{Cooling Timescales and Flare Loop Thickness \label{sec_cooling_thickness}}

The analysis described in \S \ref{sec_loop_length} only provides the loop length. The ratio between the loop's cross-sectional radius and the loop length ($\beta = r/L$), which indicates the thickness of the loop, remains to be determined. For the cool ($T \sim 1$~MK) solar active region coronal loops spatially resolved by the Transition Region and Coronal Explorer \citep[{\it TRACE},][]{Handy99}, $\beta$ ranges from 0.01 to 0.06 for loops with a half-length of $L<100$~Mm, and $\beta$ falls in the narrow range $0.007-0.02$ for larger loops \citep[Table~3 and \S 4.2 in][]{Aschwanden00}\footnote{The loop thickness here is estimated as $\beta = w / 2 / L$, where $w$ is the width of a bundle of loop threads and not an individual thread, and $L$ is the loop half-length; $w$ and $L$ measurements are taken directly from Table~3 of \citet{Aschwanden00}.}. For giant coronal flaring structures found in ONC PMS stars, \citet{Favata05} and G08a assumed $\beta = 0.1$. This assumption might not hold \citep[section 3.2 in][]{Favata05}. To check the applicability of the Reale method to big PMS flares, \citet{Favata05} performed detailed hydrodynamic simulations of the big flare in COUP star \# 1343. Their loop simulations with $\beta = 0.02$ were in good agreement with the data.

Similar to the DQ Tau flare with no (Model~1, $\zeta = 1.5$) or mild (Model~2, $\zeta = 1.1$) sustained heating, the COUP~1343 flare loop is freely decaying with no sustained heating ($\zeta \gtrsim 1.5$; G08a). Assuming similarity in flare decay cooling processes between the DQ~Tau and COUP~1343 flares and using the COUP~1343 flare as a testbed, further comparison of the cooling timescales with the light curve decay timescales for both flares allows estimation of the thickness of the DQ~Tau loop. 

The dominant cooling mechanisms of a coronal plasma are thermal conduction and radiation with timescales of \citep[e.g.][]{Cargill95}:

\begin{equation}
\tau_{con} = \frac{3 n_e k_b T L^2}{\kappa_0 T^{7/2}}, \tau_{rad} = \frac{3 n_e k_b T}{n_e^2 P(T)}
\label{eqn_cooling_times1}
\end{equation}

where $k_b$ is the Boltzmann constant, $n_e$ is the electron density, $T$ is the measured plasma temperature, $L$ is the loop length, $\kappa_0$ is the coefficient of thermal conductivity \citep{Spitzer65}, taken here as $\kappa_0 = 10^{-6}$~ergs~s$^{-1}$~cm$^{-1}$~K$^{-7/2}$, and $P(T)$ is the radiative loss function for an optically thin plasma \citep[e.g.][]{Rosner78,Mewe85}. These timescales can be expressed as functions of independent quantities already derived in our previous analyses, such as, the loop length $L$ (\S \ref{sec_loop_length}), temperature $T$, emission measure $EM$, X-ray luminosity in the wide $0.01-50$~keV band $L_{X,0.01\_50}$ (Table~\ref{tbl_trs}), as well as the quantity of interest here, the loop thickness $\beta$. At any given time $t$:

\begin{equation}
V = 2 \pi \beta^2 L^3
\label{eqn_volume}
\end{equation}

\begin{equation}
n_e(t) \simeq \sqrt{\frac{EM(t)}{V}}
\label{eqn_density}
\end{equation}

\begin{equation}
L_{X,0.01\_50}(t) = n_e(t)^2 V P(T)
\label{eqn_luminosity}
\end{equation}

where $V$ is the loop volume, so that

\begin{equation}
\tau_{con}(t) = \frac{3 k_b}{\kappa_0 T(t)^{5/2} \beta} \sqrt{\frac{EM(t) L}{2 \pi}}, \tau_{rad}(t) = \frac{3 k_b T(t) \beta}{L_{X,0.01\_50}(t)} \sqrt{EM(t) 2 \pi L^3}.
\label{eqn_cooling_times2}
\end{equation}

\begin{figure}
\centering
\includegraphics[angle=0.,width=6.5in]{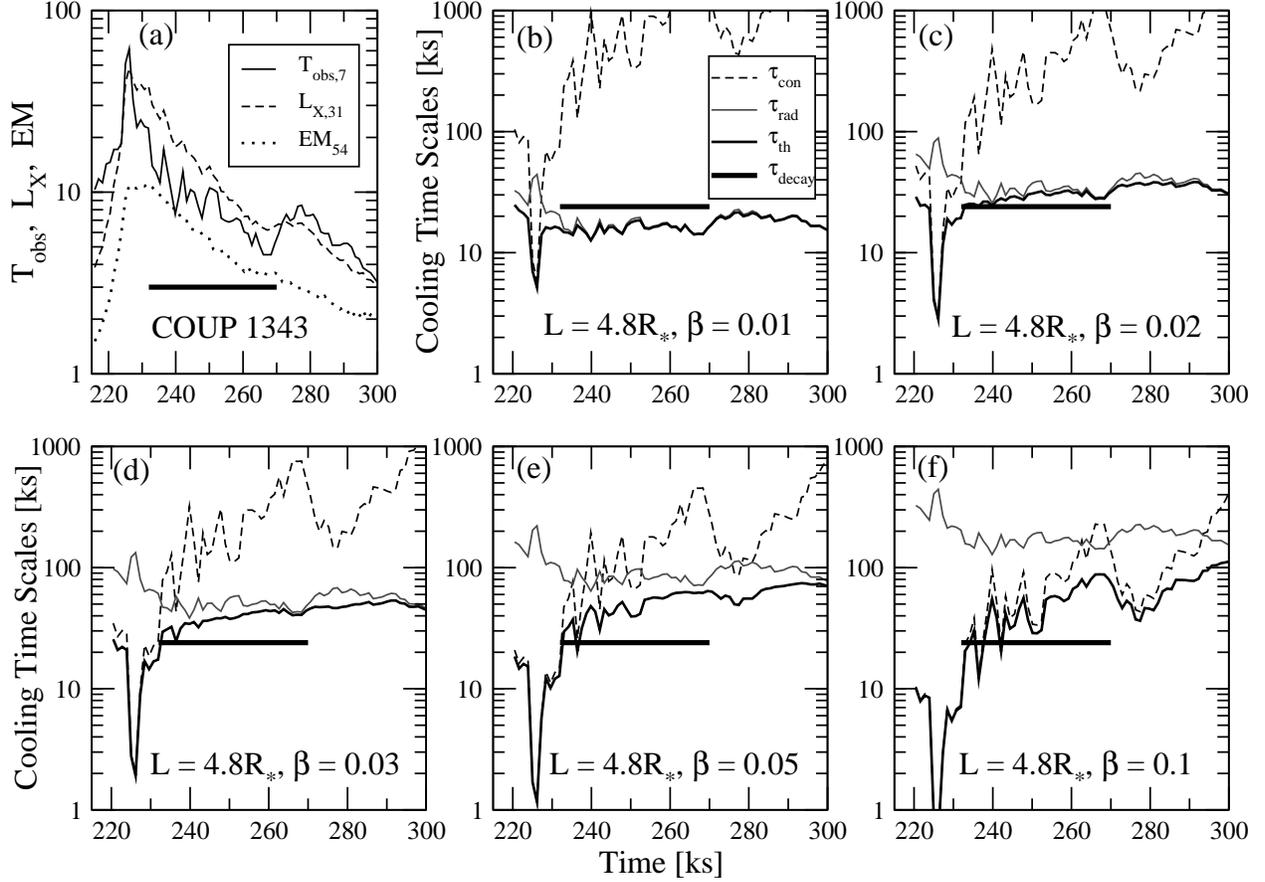}
\caption{Comparison of the derived flare cooling timescales with the observed light curve decay timescale for a testbed X-ray flare, specifically the superflare from COUP source \#~1343. (a) Evolution of the inferred plasma temperature (solid line), emission measure (dotted lines), and absorption-corrected X-ray luminosity (dashed line). The flare decay phase of interest is indicated by the thick line. (b)-(f) Evolution of the derived $e$-folding cooling time scales: thermal conduction cooling ($\tau_{con}$, thin dashed line), radiation cooling ($\tau_{rad}$, thin solid gray line), combined conduction and radiation cooling $\tau_{th}$ ($1/\tau_{th} = 1/\tau_{con} + 1/\tau_{rad}$, thin solid black line). The $e$-folding decay timescale of the observed $Chandra$ light curve ($\tau_{decay}$) is indicated by the thick solid line. Cooling time scales are derived assuming a flaring loop length of $L = 4.8$~R$_{\star}$ and five different values for the ratio of the loop cross-sectional radius
to the loop length $\beta$. \label{fig_coup1343_coolingtimes}} 
\end{figure}

For the COUP~1343 flare, the evolution of its $T$, $EM$, and $L_{X,0.01\_50}$ as derived by G08a is shown in Figure \ref{fig_coup1343_coolingtimes}a. Figure \ref{fig_coup1343_coolingtimes}b-f compares cooling timescales obtained from equation (\ref{eqn_cooling_times2}) for the loop length of $4.8$~R$_{\star}$ \citep{Favata05,Getman08a} and a range of $\beta$ with the observed light curve decay timescale of $\tau_{decay} = 24$~ks (G08a). For solar and stellar flares, if sustained heating is mild or absent, it is anticipated that from the end of the heating phase (indicated by the temperature peak) to the time of peak density (the emission measure peak) the plasma cooling is governed by conduction; at the density peak the radiation cooling and conduction times become equal; afterwards the radiation cooling dominates \citep[e.g.][]{Reale07}. During the flare decay the combined conduction and radiation cooling time $\tau_{th}$ ($1/\tau_{th} = 1/\tau_{con} + 1/\tau_{rad}$) is naturally expected to be somewhat close to the observed flare decay time $\tau_{decay}$. That is precisely what happens in the evolution of the COUP~1343 cooling timescales with a choice of $\beta = 0.02$ (panel c). In this case, during the flare decay the $\tau_{th}$ differs from $\tau_{decay}$ by no more than a factor 1.3, with the conduction and radiation times matching at the emission measure peak $t = 230$~ks and with radiation losses dominating afterwards. With a choice of $\beta < 0.02$,  $\tau_{th}$ is systematically lower than $\tau_{decay}$ (panel b), while with $\beta > 0.02$, $\tau_{th}$ is systematically higher than $\tau_{decay}$ (panels d-f) with radiation losses becoming negligible at $\beta \sim 0.1$ (panel f). For the COUP~1343 flare our cooling timescale analysis thus gives $\beta = 0.02$ and is in complete agreement with the results of the detailed simulations of \citet{Favata05}. This validates our use of cooling timescales for the estimation of the loop thickness.

\begin{figure}
\centering
\includegraphics[angle=0.,width=5.0in]{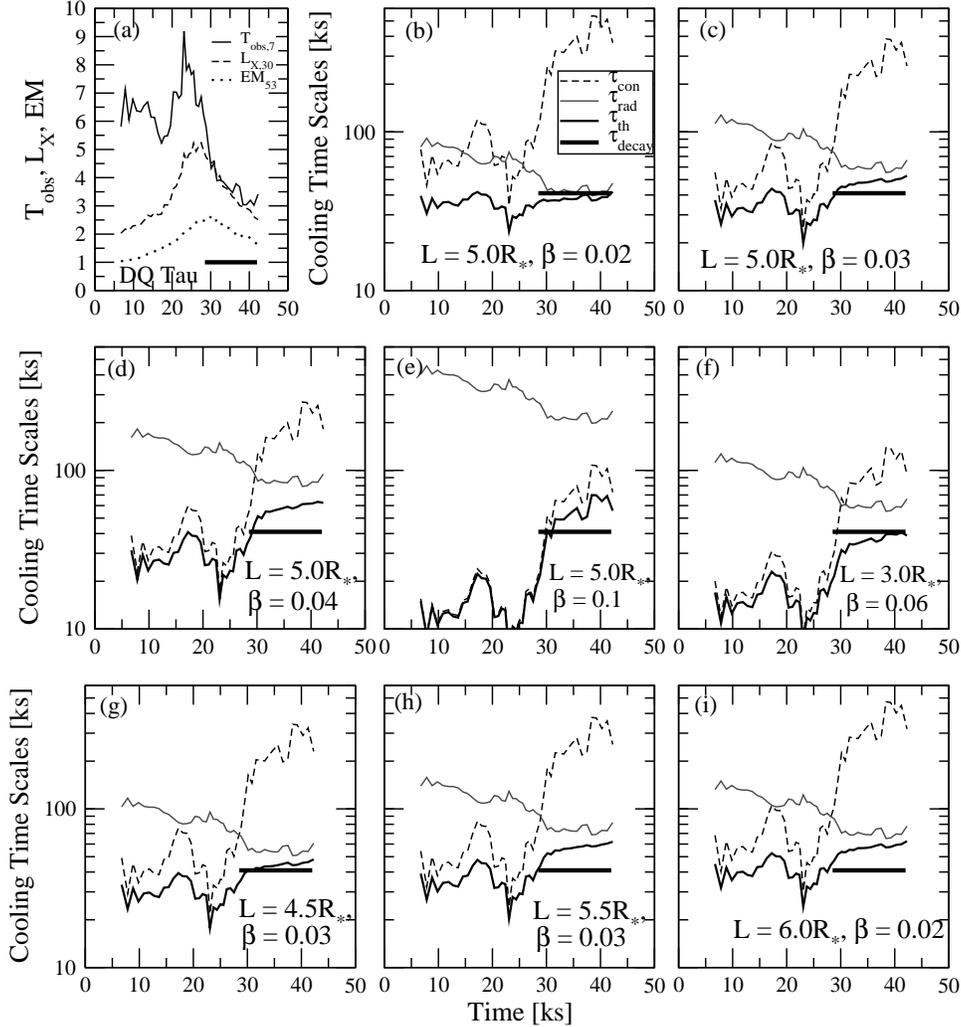}
\caption{{\small Comparison of the derived flare cooling timescales with the observed light curve decay timescale for the $Chandra$ flare toward DQ Tau using low temperature plasma component Model~1. (a) Evolution of the inferred plasma temperature (solid line), emission measure (dotted lines), and absorption-corrected X-ray luminosity (dashed line). The flare decay phase of interest is indicated by the thick line. (b)-(i) Evolution of the derived $e$-folding cooling time scales: thermal conduction cooling ($\tau_{con}$, thin dashed line), radiation cooling ($\tau_{rad}$, thin solid gray line), combined conduction and radiation cooling $\tau_{th}$ ($1/\tau_{th} = 1/\tau_{con} + 1/\tau_{rad}$, thin solid black line). The $e$-folding decay timescale of the observed $Chandra$ light curve is indicated by the thick solid line. Panels (b)-(e) present cooling timescales derived assuming the loop length of $L=5$~R$_{\star}$ for four different values of $\beta$. Panel (c) here shows the timescale pattern reminiscent of that seen in COUP \#~1343 flare (Figure \ref{fig_coup1343_coolingtimes}c). Panels (f)-(i) exemplify evolution of cooling timescales for loop lengths both less than and higher than an $L=5$~R$_{\star}$, with the values of the loop thickness parameter $\beta$ selected to emphasize cases that most closely follow the cooling timescale evolution pattern seen in panel (c).} \label{fig_dqtau_coolingtimes_model1}} 
\end{figure}

This method has also been applied to the DQ~Tau flare; the results of the modeling with the Model~1 are shown in Figure \ref{fig_dqtau_coolingtimes_model1}.  Panels b-e compare cooling timescales obtained from equation (\ref{eqn_cooling_times2}) for a loop length of $5$~R$_{\star}$ (\S \ref{sec_loop_length}) and a range of $\beta$ with the observed light curve decay timescale of $\tau_{decay} = 40.9$~ks (\S \ref{sec_xray_reduction}). With a choice of $\beta = 0.03$ (Figure \ref{fig_dqtau_coolingtimes_model1}c) the evolution of cooling timescales in DQ~Tau follows the correct pattern seen in Figure \ref{fig_coup1343_coolingtimes}c for the COUP~1343 flare. With a choice of $\beta < 0.03$, $\tau_{th}$ is systematically lower than $\tau_{decay}$ (Figure \ref{fig_dqtau_coolingtimes_model1}b), while for $\beta > 0.03$, $\tau_{th}$ is systematically higher than $\tau_{decay}$ (Figure \ref{fig_dqtau_coolingtimes_model1}d-e). The results of the cooling timescale analysis are also in agreement with the loop length analysis (\S \ref{sec_flare_analysis}): with a choice of loop length $L \ll 5$~R$_{\star}$, $\tau_{th}$ is systematically lower than $\tau_{decay}$ (Figure \ref{fig_dqtau_coolingtimes_model1}f), while for $L \gtrsim 5.5$~R$_{\star}$, $\tau_{th}$ is systematically higher than $\tau_{decay}$ (Figure \ref{fig_dqtau_coolingtimes_model1}h-i). The model $L = 4.5$~R$_{\star}$, $\beta = 0.03$ is also plausible (Figure \ref{fig_dqtau_coolingtimes_model1}g). For the DQ Tau flare our cooling timescale analysis thus gives $\beta = 0.03$ in the case of Model~1 (Figure \ref{fig_dqtau_coolingtimes_model1}), and $\beta = 0.06$ in the case of Model~2 (results are not shown). The Model~2 assumption gives a shorter and thicker flaring loop with the twice the volume of a Model~1 loop.

\section{Comparison of the $Chandra$ X-ray and IRAM mm Light Curves \label{sec_chandra_iram}}

For most of the $Chandra$ observing time the mm flux of DQ~Tau was monitored by a number of mm-band interferometers: IRAM, CARMA, and SMA. The mm data are presented by \citet{Salter10}. Figure \ref{fig_chandra_vs_iram}a compares our X-ray data with the 90~GHz ($3.3$~mm) IRAM data. The IRAM data suggest the presence of at least two microwave flares.  The end of the captured decay portion of the first microwave flare is close in time to the peak of the X-ray flare. Such a light curve behavior is consistent with the Neupert effect \citep[e.g.][]{Veronig02}. 

More precisely, the Neupert effect is a correlation between the time-integrated microwave (or hard X-ray) light curve and the rising portion of the soft X-ray light curve \citep{Neupert68}. The effect has been observed in many solar \citep[e.g.][]{Dennis93} and some stellar flares \citep[e.g.][]{Gudel02}, and can be interpreted by the classical non-thermal thick-target model \citep{Brown71,Lin76}. In this model, electrons are accelerated to high energies and spiral downwards along the magnetic field lines of the coronal loop. The high-pitch angle (high energy) electrons are trapped in the coronal region of the loop, with the highest-energy population emitting microwaves \citep[e.g.][]{Dennis88}. The low-pitch angle electrons propagate to the chromosphere. When they reach the chromosphere, they produce non-thermal hard X-rays and drive chromospheric evaporation, which fills the loop with hot plasma emitting in soft ($Chandra$ band) thermal X-rays.

\subsection{Modeling of Thermal Energy Profiles \label{sec_energy_profiles}}

Quantitative evaluation of the relation between the IRAM and $Chandra$ data of DQ~Tau can be deduced within the framework of the generalized Neupert effect introduced in \citet{Gudel96}. According to this formalism, the temporal profile of the total thermal energy $E(t)$ of the plasma in the flaring loop can be expressed as the convolution of the kinetic energy influx by non-thermal electrons into the chromosphere (assumed proportional to the observed radio flux) with radiative and conductive cooling (assumed to follow an exponential decay)

\begin{equation}
E(t) = \alpha \int_{t_0}^{t} F_R(t^{\prime}) e^{-(t-t^{\prime})/\bar{\tau}_{th}} dt^{\prime},
\label{eqn_generalized_neupert}
\end{equation}

where $\alpha$ is an assumed constant conversion factor between the thermal energy flux from electrons injected into the chromosphere and the observed radio flux $F_R$, $\bar{\tau}_{th}$ is the time-averaged cooling time of the thermal plasma in the loop, $t_0$ is the start time of the radio flare, and $t^{\prime} < t$.

\begin{figure}
\centering
\includegraphics[angle=0.,width=4.5in]{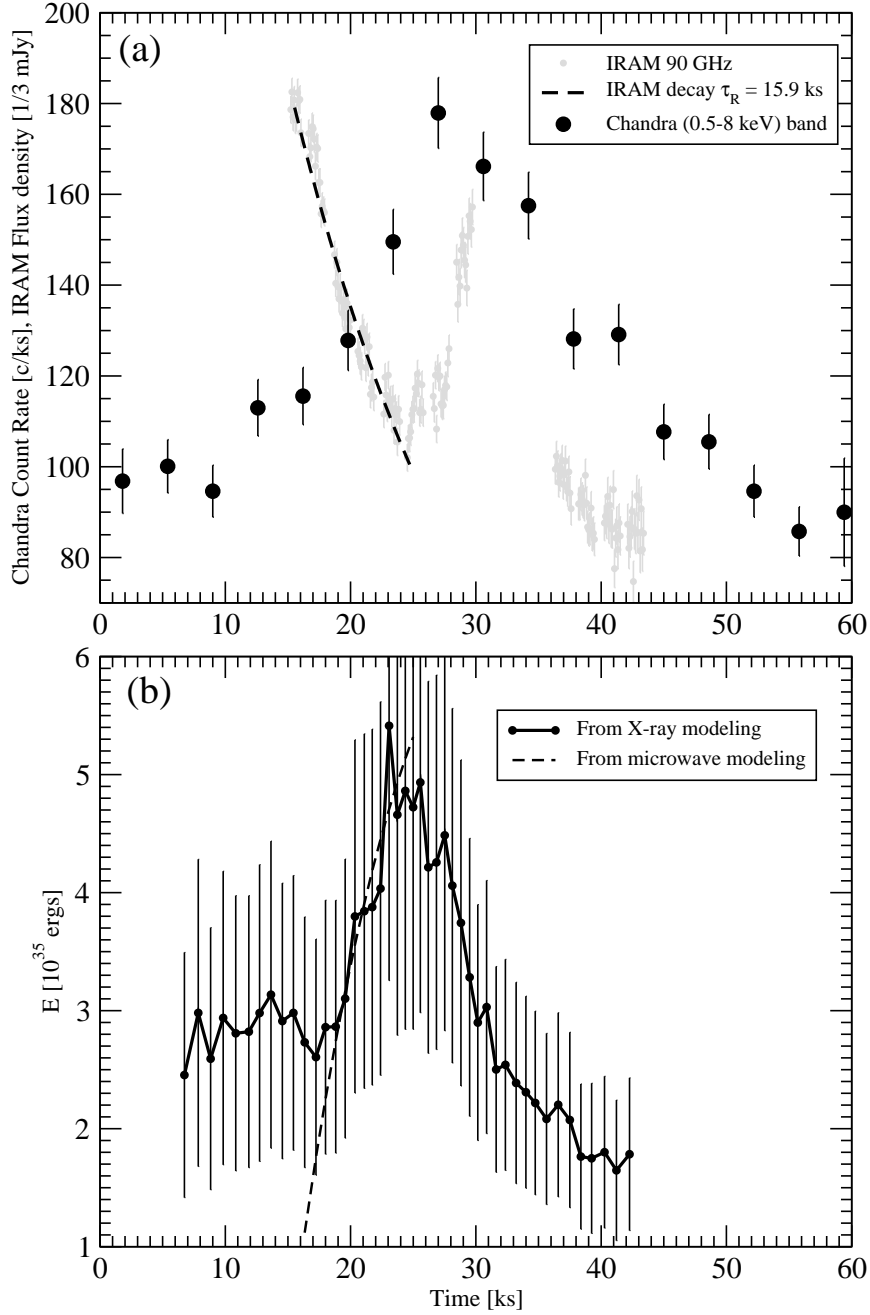}
\caption{{\small (a) Comparison between the IRAM mm (small gray circles) and $Chandra$ X-ray (big black circles) light curves for DQ Tau. The mm flare decay e-folding timescale is $\tau_R = 15.9$~ks (dashed line). (b) Indication of a Neupert-like effect. Plasma thermal energy (thick solid line) inferred from our X-ray modeling (using Model~1) compared to the convolved IRAM light curve using formula (12) with parameters $C_1 = 0.5 \times 10^{35}$~ergs and $\alpha \times F_R(t_1) = 0.8 \times 10^{32}$~ergs~s$^{-1}$  (thick dashed line). In the X-ray modeling, the energy points are derived from overlapping time segments (\S \ref{sec_trs}); thus the energy errors are not independent.} \label{fig_chandra_vs_iram}} 
\end{figure}

In the case of DQ~Tau, the captured late decay phase of the mm flare observed with IRAM can be fit to an exponential model

\begin{equation}
 F_R(t) = F_R(t_1) e^{(t1-t)/\tau_R},
\label{eqn_iram_flux}
\end{equation}

where $t_1  = 15.5$~ks is the earliest time of the portion of the mm flare decay that was observed and $\tau_R = 15.9$~ks is the decay e-folding timescale during that phase (Figure \ref{fig_chandra_vs_iram}a). The expected thermal energy profile is then

\begin{equation}
E(t) = C_1 + \alpha F_R(t_1) \int_{t_1}^{t} e^{(t1-t^{\prime})/\tau_R} e^{-(t-t^{\prime})/\bar{\tau}_{th}} dt^{\prime},
\label{eqn_generalized_neupert_DQTau}
\end{equation}

where the averaged plasma cooling time over the rise phase of the $Chandra$ flare is $\bar{\tau}_{th} \sim 30$~ks (see Figure \ref{fig_dqtau_coolingtimes_model1}c) and the $C_1$ is a constant term accounting for the integration over the unseen part of the mm flare. The integral in equation (12) can be solved analytically; however two constant parameters remain unknown: the expected total thermal energy at the end of the unseen part of the mm flare ($C_1$) and the rate of the injected energy at the beginning of the captured decay of the mm flare ($\alpha \times F_R(t_1)$).

Equation (12) defines the thermal energy profile using the IRAM data. On the other hand, using equations (5) and (6) the thermal energy $E(t) = 3 n_e(t) k_b T(t) V = 3 k_b T(t) \sqrt{2 \pi EM(t)} \beta L^{3/2}$ can be inferred from our X-ray flare analysis; the resulted profile (using X-ray Model~1) is shown as the thick solid curve in Figure \ref{fig_chandra_vs_iram}b. Although the energy is subject to significant uncertainty, we are not so much interested in the individual values as in the general evolution of thermal energy that is observed: an early rise phase ($t < 19$~ks), when the energy is approximately constant around $3 \times 10^{35}$~ergs; a late rise phase, when the energy rapidly rises to $\sim 5 \times 10^{35}$~ergs within a 6~ks period; and a flare decay phase, when the energy falls to $\sim 2 \times 10^{35}$~ergs during the remaining 15~ks period. For Model~2 the evolution is similar, but the energies are systematically higher by factors of $\sim 1.2-1.4$.

A family of solutions to equation (12) exists with constant parameter values around $C_1 = 0.5 \times 10^{35}$~ergs and $\alpha \times F_R(t_1) = 0.8 \times 10^{32}$~ergs~s$^{-1}$, resulting in thermal energy profiles (such as the dashed curve in Figure \ref{fig_chandra_vs_iram}b) that resemble the profile derived from the X-ray modeling during the late, but not the early, stage of the X-ray flare rise phase\footnote{Parameters $C_1 = 0.5 \times 10^{35}$~ergs and $\alpha \times F_R(t_1) = 0.8 \times 10^{32}$~ergs~s$^{-1}$ are the values to use in equation (12) to match the thermal energy profile from the X-ray modeling with the Model~1. For matching the energy profile from the X-ray modeling with Model~2, the best values to use are $C_1 = 2 \times 10^{35}$~ergs and $\alpha \times F_R(t_1) = 0.6 \times 10^{32}$~ergs~s$^{-1}$.}. This resemblance of the thermal energy profiles derived independently from the microwave and X-ray modeling provides a quantitative indication of a Neupert-like effect. The divergence of the dashed and solid curves in Figure \ref{fig_chandra_vs_iram}b for the very early rise phase of the X-ray flare ($t < 19$~ks) could be due to three possible effects that we have not modeled: 1. X-ray emission excess from flaring events in other coronal loops; 2. time dependence of the energy conversion factor $\alpha$ during the early decay phase of the microwave flare; 3. inaccurate estimation of the conduction cooling timescale $\tau_{con}$, electron density $n_e$, and thermal energy $E$ during the very early stages of chromospheric evaporation.

Finally, we assume that the second mm flare detected by IRAM is related to a separate and independent flaring event in another coronal loop. The weaker X-ray emission associated with the secondary event has a negligible impact on the observed decay phase of the primary large X-ray flare.

\section{Comparison with COUP PMS and Flares on Older Stars \label{sec_coup_star_comparison}}

\begin{figure}
\centering
\includegraphics[angle=0.,width=6.5in]{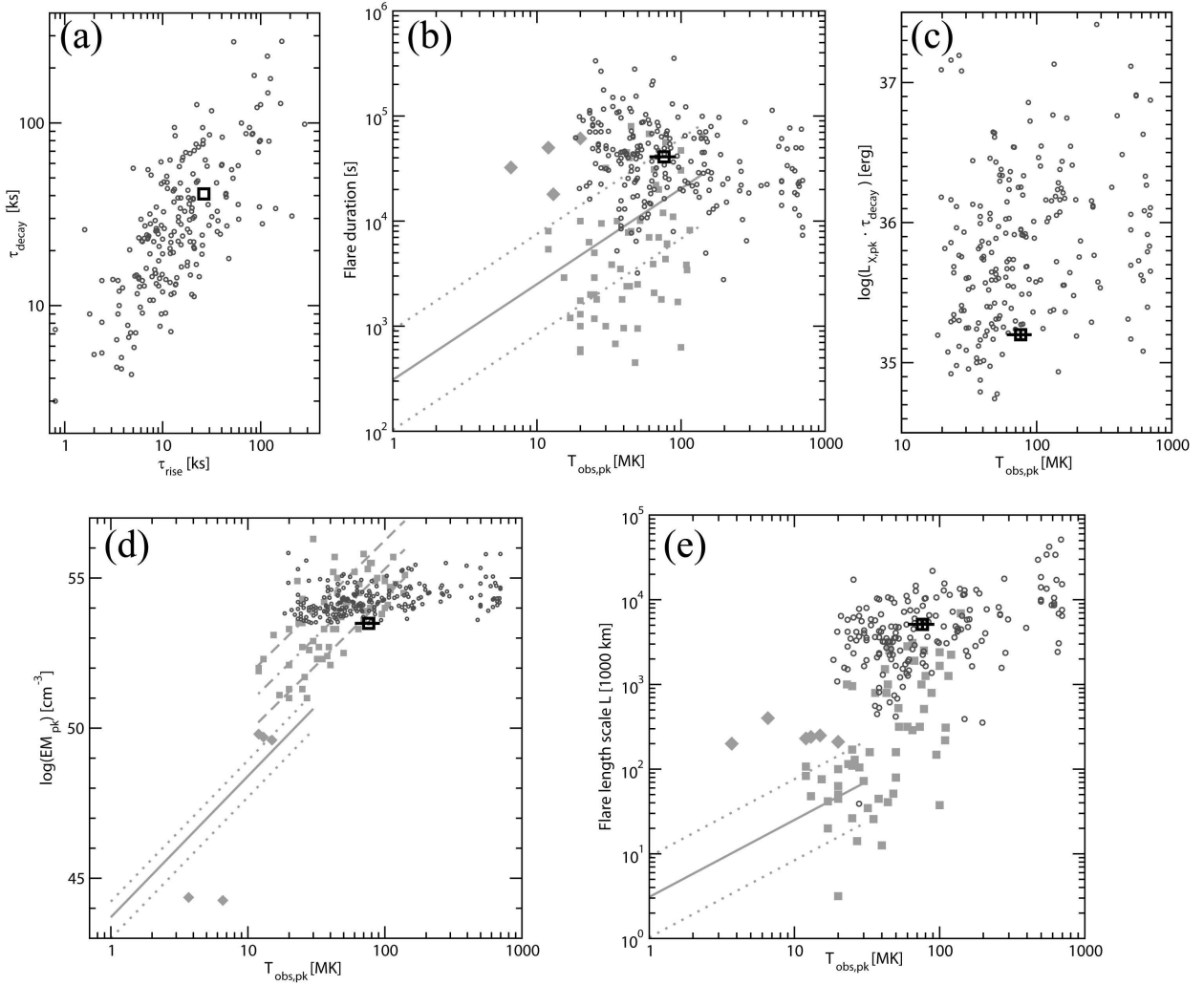}
\caption{Comparison of the DQ Tau flare properties with those of the COUP PMS stars (panels (a)-(e)) and flares on older stars (panels (b), (d), (e)). (a) Flare decay vs.\ rise timescales. (b) Flare duration vs. peak flare temperature. (c) Time-integrated flare energy (determined from the flare peak X-ray luminosity times the flare decay timescale) vs. peak flare temperature. (d) Peak flare emission measure vs. peak flare temperature. (e) Flare loop length vs. peak flare temperature. Large flares from the COUP sample are shown as black circles. In panels (b), (d), and (e), the stellar flares compiled by \citet{Gudel04} are marked by gray boxes with their linear regression fit and 1$\sigma$ range (in panel (d)) shown as gray dashed-dotted and dashed lines, respectively. Regression fit and 1$\sigma$ range for solar-stellar (panel (b)) or just solar (panels (d) and (e)) flares compiled by \citet{Aschwanden08} are marked by gray solid and dotted lines with representative solar LDEs (compiled by G08a) discussed in the text shown as gray diamonds. The DQ Tau flare is marked by the thick black square with bars representing systematic differences between the flare properties derived using modelling with the low temperature plasma component Model~1 and Model~2. \label{fig_comparison_with_coup_oldstars}} 
\end{figure}

Figure \ref{fig_comparison_with_coup_oldstars} compares the DQ~Tau X-ray flare properties (indicated with a black box) with those of the COUP sample of flares analyzed by G08a (circles), stellar flares from PMS and older active stars compiled in \citet{Gudel04} (gray boxes), and solar-stellar trends obtained by \citet{Aschwanden08} (solid and dotted gray lines). It is important to note that the analysis of G08a is limited to COUP sources with $>4000$ counts and to flares with a peak count rate that is $\ga 4$ times the characteristic level. Thus, in Figure~\ref{fig_comparison_with_coup_oldstars} the lower boundaries of the COUP flare duration, energy, emission measure, and loop size correspond to the imposed selection. Panel (a) shows that the rise and decay e-folding timescales of the DQ~Tau flare lie well within the locus of the COUP sample of flares, indicating similar flare morphology. Panels (b) and (e) show that the duration, peak plasma temperature, and length of the associated coronal structure of the DQ~Tau flare are typical of the COUP sample of flares, and are consistent with only the longest and hottest flares from older active stars. However, at a given temperature, the DQ~Tau flare emission measure is at the lower boundary of the COUP and stellar loci (panel (d)). As a result of the relatively low emission measure, the DQ~Tau flare energy is only comparable to that of the least energetic flares from the COUP sample (panel (c))\footnote{In panel (c) the DQ~Tau flare energy estimate is given for the $(0.5-8)$~keV energy band to be compatible with the COUP data.}. Compared to typical solar flares, the COUP/DQ~Tau flares are $10-100$ times longer lasting and hotter, and are associated with $100-1000$ times larger coronal structures. 

The $L$ vs.\ $T_{obs,pk}$ information in panel (e) must be interpreted with care. Since the locus of typical solar flares (gray line) represents direct observational measurements of flare length scales, and the plotted symbols are model-derived estimates, it is prudent to suspect that the relatively large loops inferred for ONC PMS stars (black circles) may represent systematical biases introduced by single-loop modeling. However, about half of the 20 flares from older active stars (gray boxes) that lie within the locus of solar flares ($T_{obs,pk} \la 40$~MK) have been modeled using the same single-loop approach of R97; no large upward bias in their inferred loop lengths is seen. We caution against concluding that PMS flares follow a different $L$ vs.\ $T_{obs,pk}$ correlation than solar flares (grey line, extended), because observational selection effects could easily be hiding a very large population of flares that do follow the extended solar locus.

The difference between DQ~Tau and typical COUP flare emission measures can be attributed either to the COUP data selection effect mentioned above, or to the differences in flare physical processes. In the latter case, with similar flare morphologies, durations, temperatures, and loop lengths, the emission measure difference can be explained via distinct coronal loop geometries (e.g., a DQ Tau loop that is thinner than typical loops associated with flares from the COUP sample) and/or by distinct efficiencies in chromospheric evaporation that in turn can be linked to differences in the magnetic reconnection processes (\S \ref{sec_origin}).

\section{Discussion \label{sec_discussion_conclusions}}

\subsection{Single Loop versus Multiple Loop Scenarios \label{sec_single_vs_multiple}}

Large flares on the Sun often involve arcades of dozens or hundreds of sequentially reconnected magnetic loops, which are often observed to have different temperatures with the cooler loops lying below the hotter ones \citep[e.g.][and references therein]{Reeves02}. Despite the morphological complexity of such flares, their X-ray light curves often have simple exponential rise and decay shapes.\footnote{Images of complex solar flaring loop arcades spatially resolved by {\it TRACE} \citep[][]{Handy99} at Extreme Ultraviolet (EUV) wavelengths and their associated X-ray light curves obtained by the Geostationary Operational Environmental Satellites \citep[{\it GOES},][and references therein]{Garcia94} can be found and inspected at the {\it TRACE} flare catalog web site {\url{http://hea-www.harvard.edu/trace/flare\_catalog/index.html}} (see also Figure~\ref{fig_solar_flares}).} In view of this solar analogy, the concept of the single-loop approach and its application to powerful spatially unresolved stellar flares might be questionable.

R97 have applied their single-loop method to flare decays of 20 solar M- and C-class flares monitored by the Soft X-ray Telescope \citep[SXT,][]{Tsuneta91} on board the {\it Yohkoh} solar observatory satellite \citep{Ogawara91}. Loop sizes derived using the method were shown to agree with the length of X-ray structures measured from direct inspection of SXT images (Figure~6 in R97). In Appendix~\ref{sec_solar_flares} we apply the approach of R97 to 5 solar X-class flares that are clearly associated with arcades of multiple loops seen in EUV {\it TRACE} images. In contrast to the general view that single loop models tend to overestimate flaring loop sizes of complex flare events, for all of our 5 testbed flares the equations of R97 yield a single-loop length comparable to or shorter than the lengths of the individual EUV loops measured from the {\it TRACE} images. 

Here we are unable to give a definite answer to the question of why the loop lengths of multi-loop X-class flares derived using the R97 approach can be comparable to the observed loop lengths, considering that the application of the single-loop approach to multi-loop flares is a priori incorrect.  

The consistency may be by chance. For instance, for all 5 flares, the inferred slope $\zeta$ in the $\log T$~--~$\log \sqrt{EM}$ diagram is found to be close to $\sim 0.4$. Let us ignore for a moment the fact that the physical meaning of the slope $\zeta$ for multi-loop flare arcades is generally different from that of a single loop (i.e. different plasmas in different loops heated and cooled sequentially versus a single plasma in a single loop heated and cooled). Within the framework of the single-loop model of R97, $\zeta \sim 0.4$ is close to the lowest values allowed, and the equations of R97 produce smaller loop lengths at smaller slopes. In fact, in the regime of low $\zeta$ ($\zeta < 0.7$), the loop length is steeply decreasing with decreasing $\zeta$, by a factor of 10 when $\zeta$ changes from 0.7 to 0.4. 

Contrary to the ``by chance'' explanation above, loop length numbers can be comparable due to the presence of a single loop (or localized multiple loops ignited simultaneously and undergoing similar heating and cooling processes) that dominates the flare X-ray emission. For instance, an arcade-like structure with a single primary loop dominating the rise and early decay phases of a flare was proposed (based on a detailed modeling) for the complex flare on Proxima Centauri \citep{Reale04}. At least for the best studied of the 5 solar X-class flares considered here, the Bastille Day flare, we do no find compelling reasons for the presence of such a dominant X-ray emitting structure (Appendix~\ref{sec_solar_flares}). 

Clearly, even if the loop lengths inferred via the single-loop method happen to be comparable to the observed values for such complex flare arcade events, the method would not predict a correct loop geometry and would likely underestimate the emitting volume by at least a factor comparable to the number of instantaneously heated loops in the arcade, $N_{loop}$, and would overestimate the average electron density of emitting plasma by $\sim \sqrt{N_{loop}}$. For instance, during the Bastille Day flare arcade, $N_{loop} \sim 20$ out of the $\sim 100$ observed loops fired near the flare peak time \citep{Aschwanden01}.

For all 5 solar X-class flares considered here, the inferred slope $\zeta$ in the $\log T$~--~$\log \sqrt{EM}$ diagram is found to be close to the value of $\sim 0.4$. Observationally, such a value is among the lowest values seen in solar flares \citep[e.g.][]{Sylwester93}. Such a shallow slope indicates continued heating during the decay phase of the X-ray emission. Continued heating is also what one would expect for multi-loop two-ribbon flares where the decay phase could be entirely driven by the heating released through sequential reconnection in individual loops \citep{Kopp84}.

The slope $\zeta \ga 1$ in the $\log T$~--~$\log \sqrt{EM}$ diagram for the DQ~Tau flare is significantly larger than the $\zeta \sim 0.4$ found for complex powerful solar flares (Appendix~\ref{sec_solar_flares}), suggesting heating behaviour different from the 5 solar X-class flares. The possibility that the DQ~Tau flare could involve an arcade-like structure with a single loop dominating the rise and early decay phases, as was proposed for the complex flare on Proxima Centauri \citep{Reale04}, can not be excluded. On the other hand, within the framework of a model involving explosive chromospheric evaporation into a single loop \citep[e.g.][]{Reale07}, an evaporation time of only $3-5$~ks is needed for a plasma with temperature $T_{obs} \sim 60-90$~MK (\S \ref{sec_trs}) to propagate with an isothermal sound speed $v_s$$\rm{[cm~s^{-1}]}$$ = \sqrt{(5/3 k_b T_{obs})/(\mu m_p)} = 1.5 \times 10^4 \sqrt(T_{obs}\rm{[K]})$ \citep[e.g.][]{Aschwanden00} and to fill a DQ Tau flaring loop of a semi-length $L=4-5$~R$_{\star}$. The observed rise timescale of the DQ Tau light curve, $\tau_{rise} = 26.3$~ks, is much longer than $3-5$~ks and thus might indicate a complex history of heating during the early phase of the flare, e.g., due to multiple loops.

It is also worth mentioning that the time delay between temperature and emission measure peaks observed in the DQ Tau flare (\S \ref{sec_trs}) can not be used to argue in favor of a single loop event. This effect is commonly seen in multi-loop flares and can be explained by invoking the principle of linear superposition (Appendix~\ref{sec_solar_flares}). Likewise, the Neupert-like effect observed in the DQ Tau flare (\S \ref{sec_chandra_iram}) can not be used to argue in favor of a single loop event. If the Neupert effect occurred in each loop of a multi-loop system, then an observation that did not spatially resolve the loops would show the Neupert effect. Low-resolution observations of multi-loop solar flares commonly exhibit the Neupert effect \citep{Veronig02}.
  
After all, in view of the fact that the observed timescales, temperatures, and X-ray luminosities of PMS stars (DQ~Tau and  ONC stars) are much higher than those of typical solar flares (\S \ref{sec_coup_star_comparison}), one could argue that the solar analogy might not be directly applicable. PMS and magnetically active older stars possess stronger, than solar, surface and global magnetic fields and larger volumes for magnetic fields to interact, so the extreme flaring behaviour, beyond solar analogy, is expected \citep[e.g. \S 2.2 in][and \S \ref{sec_origin} in this work]{Benz10}. But then the single-loop approach developed on solar analogy might not be applicable either. 

We conclude that we can neither prove the presence of loop arcades analogous to the solar cases, nor refute the presence of single flaring loops longer than any seen on the Sun. The results from the single-loop X-ray modeling presented in \S \ref{sec_single_loop_modeling} and \S \ref{sec_energy_profiles} should not be treated as definitive and should be considered with caution.

\subsection{Origin of the DQ~Tau Flare \label{sec_origin}}

The origin of the big flares from the COUP sample themselves is unclear. For some of the 32 most powerful COUP flares whose inferred coronal structures reach several stellar radii, \citet{Favata05} suggest magnetic loops linking the stellar photosphere with the inner rim of the circumstellar disk. For the extended sample of $>200$ flares G08ab find that large (a few to several stellar radii in length based on the single-loop approach of R97) coronal structures are present in both disk-bearing and disk-free stars; however G08ab also find a subclass of super-hot flares with peak plasma temperatures exceeding 100~MK that are preferentially present in highly accreting systems. G08ab further propose that the majority of big flares from the COUP sample can be viewed as enhanced analogs of the rare solar long-decay events (LDEs; see \S 7.3 in G08b and references therein). LDEs are eruptive solar flare events that produce X-ray emitting arches and streamers with altitudes reaching up to several hundred thousand kilometers ($L \sim 0.5-1$~R$_{\odot}$). In panels (b, d, e) of Figure \ref{fig_comparison_with_coup_oldstars}, representative solar LDEs (gray diamonds; compiled in G08a) are shown to have systematically higher flare durations and coronal lengths than the more typical solar flares (solid and dotted gray loci). We wish to emphasise that the analogies of the big flares from the COUP sample to solar LDEs (``growing'' systems of giant multiple loops) are not based on information about the specific detailed geometry of COUP flares (the geometry is really un-known to us and is only simplistically modeled as a large single loop), but are instead based on the simple fact that the observed flare durations and model-inferred characteristic loop scales for COUP flares are the largest ever reported from PMS stars, provided that these model-inferred loop scales are close to the truth. The most widely accepted model for the origin of LDEs is that the impulsive flare near the solar surface ($L\la 10^{-2}$~R$_{\odot}$) blows open the overlying large-scale magnetic field with subsequent reconnection of large-scale magnetic lines through a vertical current sheet. The large-scale magnetic field of PMS stars is likely far stronger than in the Sun, and can sustain giant X-ray arches and streamers with sizes $L\sim 1-10$~R$_{\star}$. For the DQ~Tau flare a different mechanism from that of a solar LDE-like reconnection can be considered.

There are at least three independent supporting lines of evidence suggesting that the DQ~Tau flare could be produced through a process of colliding magnetospheres.
\begin{enumerate}

\item Although the DQ~Tau flare duration, morphology, and plasma temperature are typical of those of big flares from the COUP sample (\S \ref{sec_coup_star_comparison}), the probability of observing a big COUP-like flare is small. Within the COUP observation window ($\sim 1200$~ks) on average 3 big flares per bright PMS star were detected. This number is derived from the analysis of the 161 brightest X-ray PMS stars (G08a); the frequency of big flares for the remaining 1200 fainter COUP PMS stars is even lower. The DQ Tau $Chandra$ observation was designed for an X-ray flare to appear close to the orbital phase of $\Phi = 0.98$, which corresponds to the peak of the April~2008 large sub-mm flare \citep{Salter08}. The detected $Chandra$ flare indeed peaks within $15$~ks of this orbital phase. The probability of detecting a big COUP-like flare within $15$~ks of a pre-specified point in time is small ($Prob < 15 \times 2 / (1200 / 3) = 0.075$).

\item Due to the lack of systematic monitoring observations in mm-bands, mm flare activity in PMS stars has been rarely reported \citep[e.g.][]{Bower03, Salter10}. The DQ~Tau X-rays are accompanied by a relatively unique non-thermal mm activity. The X-rays and mm are likely related through a Neupert-like effect (\S \ref{sec_chandra_iram}). The observed re-appearance of this unique mm activity near periastron is proposed to be associated with processes caused by the interacting magnetospheres \citep{Salter08,Salter10}.

\item It is natural to expect large-scale magnetic structures to be involved in a process of colliding magnetospheres. The loop flaring sizes inferred from the X-ray flare analysis ($4-5$~R$_{\star}$) are comparable to half the separation of the DQ~Tau components near periastron (\S \ref{sec_loop_length}). This third item of evidence is completely model-dependent and should be considered with caution (\S \ref{sec_single_vs_multiple}).

\end{enumerate}

\subsection{Energetics and Loop Geometry \label{sec_e_l_b}}

Based on the concept of interacting magnetospheres, we can further comment on energetics and possible flare loop geometry.

We can show that a crude estimate of the power expected to be dissipated by magnetic reconnection from colliding magnetospheres in DQ~Tau is in agreement with the derived peak flare X-ray luminosity ($L_{X,0.5\_8} = (4-5) \times 10^{30}$~ergs~s$^{-1}$; see Table \ref{tbl_trs}) and the modeled rate of the kinetic energy injected into the chromosphere by non-thermal electrons ($\alpha \times F_R(t_1) = (0.6-0.8) \times 10^{32}$~ergs~s$^{-1}$; \S \ref{sec_chandra_iram}).

\begin{figure}
\centering
\includegraphics[angle=0.,width=6.4in]{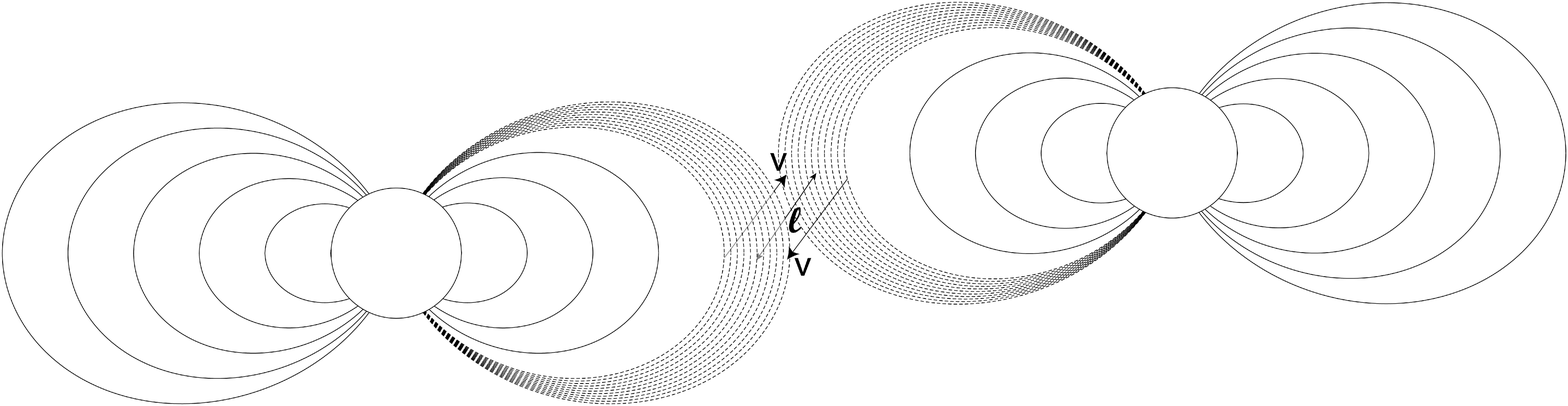}
\caption{Illustration of the process of interacting magnetospheres in DQ~Tau (in 2D projection plane). Just before periastron, with a component separation of $\sim 10$~R$_{\star}$, the magnetospheres collide with velocity $v \sim 100$~km~s$^{-1}$. Large-scale magnetic field lines of both components at $\ga 4$~R$_{\star}$ above the stellar surface violently interact (lines above $5$~R$_{\star}$ are ignored here). On the plane perpendicular to the velocity vector, the interacting magnetic lines occupy an area $A \sim 6 \times R_{\star}^{2}$ (hatched; assuming dipolar topology). The total amount of magnetic energy within the interacting volume of width $l$ for both stars is $E_m \sim 2 \times (B^{2}/8\pi) \times l \times A$ and this is released over a timescale $\tau_R \sim l/v$. \label{fig_magneto_inter}} 
\end{figure}

For a large scale dipolar topology and a magnetic field $B(L) \simeq B_{ph}/(L/R_{\star}+1)^3$ with photospheric field strengths in the range $1-6$~kG consistent with measurements of Zeeman broadening and circular polarization of photospheric lines in PMS stars \citep[e.g.][]{Johns-Krull07,Donati08}, the field strength at distances of $4-5$~R$_{\star}$ from the stellar surface is expected to be $B(4-5 \rm{R}_{\star}) \simeq 5-50$~G. Within the framework of the simple magnetic reconnection model shown in Figure~\ref{fig_magneto_inter}, the initially dipolar fields of average strength $B = 5-50$~G at $4-5$~R$_{\star}$ from the stellar surface colliding at a speed of $v \sim 100$~km/s \citep{Mathieu97} reconnect with the rate of energy release $E_m/\tau_R \sim 10^{30} - 10^{32}$~ergs~s$^{-1}$, which is in agreement with the numbers given above. It is interesting to note that the analogous procedure applied to cases of star-planet magnetic interaction have difficulty explaining an excess X-ray emission associated with stellar chromospheric hot spots rotating synchronously with close-in giant planets \citep{Lanza09}.

The model of interacting magnetospheres in RS~CVn-type binaries by \citet{Uchida85} predicts the existence of X-ray emitting loop structures connecting the stars. The model of star-planet magnetic interaction by \citet{Lanza09} expects complex topologies, including loops connecting the planet with the stellar surface. While our X-ray modeling method and many related results are limited to the scenario of a single X-ray emitting loop on a single star, other scenarios for the DQ~Tau loop geometry are possible: multiple X-ray emitting loops on a single star (\S \ref{sec_single_vs_multiple}), two or more loops appearing on both stars simultaneously, or a single or multiple loops connecting the two stars. For our modeled scenario, where the loop's footprints are anchored on the surface of a single star, potential destruction of the loop by centrifugal forces is not a serious issue. The derived length for the X-ray emitting structure of $4-5$~R$_{\star}$ is only comparable to and does not exceed the Keplerian corotation radius $R_{cor} = 4.7$~R$_{\star}$\footnote{The Keplerian corotation radius where the centrifugal force balances the gravitational force is $R_{cor} = (G \times M \times P^2/4 \times \pi^2)^{1/3}$. For the DQ~Tau binary components with similar masses of $M=1.6$~M$_{\odot}$ and rotational periods of $P=3$~days, $R_{cor} = 4.7$~R$_{\star}$.}.

\section{Conclusions \label{sec_conclusion}}

The young high-eccentricity binary DQ~Tau exhibits powerful recurring mm flaring attributed to collisions between the two stellar magnetospheres \citep{Salter08,Salter10}. The separation of only $\sim 8$~R$_{\star}$ between the binary components at periastron  implies magnetospheric interactions at characteristic scales and magnetic strengths comparable to those of relatively rare big X-ray flares from single COUP PMS stars (G08ab). To search for X-ray emission arising from processes associated with colliding magnetospheres, we performed simultaneous $Chandra$ X-ray and IRAM/CARMA/SMA mm observations of an orbital segment of DQ~Tau near the peaks of the previously detected mm flares. We report here a $Chandra$ detection of a long X-ray flare accompanied by mm activity close to periastron. The mm observations are discussed in more detail in \citet{Salter10}. 

We start by addressing  light photon pile-up in the $Chandra$ data (\S \ref{sec_chandra_reduction}). Further comparison with the constant and soft X-ray emission of DQ Tau at an orbital phase away from periastron detected by {\it XMM-Newton} shows similar spectral characteristics between the time-integrated {\it XMM-Newton} spectrum and the low temperature plasma component of the time-integrated $Chandra$ spectrum (\S \ref{sec_chandravsxmm}). Derived from the single-loop flare model of R97, the half-length of the X-ray emitting structure of $L=4-5$~R$_{\star}$ is comparable to half the binary component separation at periastron (\S \ref{sec_loop_length}). Inferred from the cooling time-scale analysis, the cross-sectional radius of the loop is $r = 0.03-0.06 L$ (\S \ref{sec_cooling_thickness}). In view of a possible solar analogy, the concept of the single-loop approach and its application to DQ Tau flare might be questionable (\S \ref{sec_single_vs_multiple}). Thus, both results on the loop length and cross-sectional radius of the loop should be considered with caution. The coincidence of the X-ray flare peak with the end of one of the mm flares, and the resemblance of the thermal energy profiles derived independently from the microwave and X-ray single-loop modeling, give an indication for a Neupert-like effect (\S \ref{sec_chandra_iram}). Comparison with the large flares from the COUP sample shows that the DQ~Tau flare duration, morphology, and plasma temperature are typical of COUP flares (\S \ref{sec_coup_star_comparison}). Three lines of evidence (the low probability of detecting a COUP-like flare, the Neupert effect apparent in the relative timing of the X-ray and mm flares, and the similarity between the binary separation and the inferred loop size) support the notion that the DQ~Tau flare could be produced through a process of colliding magnetospheres (\S \ref{sec_origin}). The third evidence is fully model-dependent and should be considered with caution (\S \ref{sec_single_vs_multiple}). The inferred flare peak X-ray luminosity and the modeled rate of the kinetic energy injected into the chromosphere by non-thermal electrons are in agreement with an estimate for the power dissipated by magnetic reconnection within the framework of a simple model of interacting magnetospheres (\S \ref{sec_e_l_b}). Our flare analysis is limited to the scenario of a single X-ray emitting loop on a single star; other scenarios for the DQ~Tau loop geometry are possible: multiple X-ray emitting loops on a single star (\S \ref{sec_single_vs_multiple}), two or more loops appearing on both stars simultaneously, or a single or multiple loops connecting the two stars (\S \ref{sec_e_l_b}).

Future coordinated multi-wavelength observation campaigns of DQ~Tau, spanning a larger range of orbital phase and more wavebands than in this study, are highly desirable. These will allow further investigation of the physical processes during an interaction of magnetospheres. Due to the predictable and bi-weekly recurrence of flare activity near periastron, DQ~Tau offers an excellent opportunity to persistently study the poorly understood connection between particle acceleration and coronal heating in stellar flares.

\acknowledgements We thank Leisa Townsley (Penn State University) for her role and financial support in the development of the wide variety of ACIS data reduction and analysis techniques and tools used in this study. We thank the referee Manuel G{\"u}del (University of Vienna) for his time and many very useful comments that improved this work. We thank Lyndsay Fletcher (University of Glasgow) for useful discussions on solar flares and for her help with the {\it GOES} data analysis. We thank Fabio Reale (Universita di Palermo), Rachel Osten (Space Telescope Science Institute), Stephen Skinner (University of Colorado), Hugh Hudson (SSL/UC Berkeley), Eric Feigeslon (Penn State University), Colin Johnstone (University of St. Andrews), Moira Jardine (University of St. Andrews), Jan Forbrich (Harvard-Smithsonian CfA), Joel Kastner (Rochester Institute of Technology), Laurent Loinard (CRyA-UNAM), Eric Stempels (Uppsala University), Scott Wolk (Harvard-Smithsonian CfA), Nancy Brickhouse (Smithsonian Astrophysical Observatory), Beate Stelzer (INAF - Osservatorio Astronomico di Palermo), and Ettore Flaccomio (INAF - Osservatorio Astronomico di Palermo) for helpful discussions. We also thank the CXC staff for their help with the $Chandra$ observation scheduling. This work is supported by the $Chandra$ GO grant SAO~GO0-11029X (K. Getman, PI) and the $Chandra$ ACIS Team contract SV4-74018 (G. Garmire, PI).

\appendix
\section{Application Of The Single Loop Model To Solar X-class Flares \label{sec_solar_flares}}

R97 applied their single loop approach to solar X-ray flares observed with {\it Yohkoh/SXT} and showed that the method provided loop scaling sizes comparable to those measured in SXT images. In this section we further check on the validity of the method by applying it to solar X-class flares clearly associated with multiple loop arcade geometries spatially resolved by {\it TRACE}. To better mimic stellar X-ray observations, which cannot resolve flare structures, the analysis of these solar X-class flares is carried out on disk-integrated X-ray data obtained by {\it GOES}.

\begin{figure}
\centering
\includegraphics[angle=0.,width=6.4in]{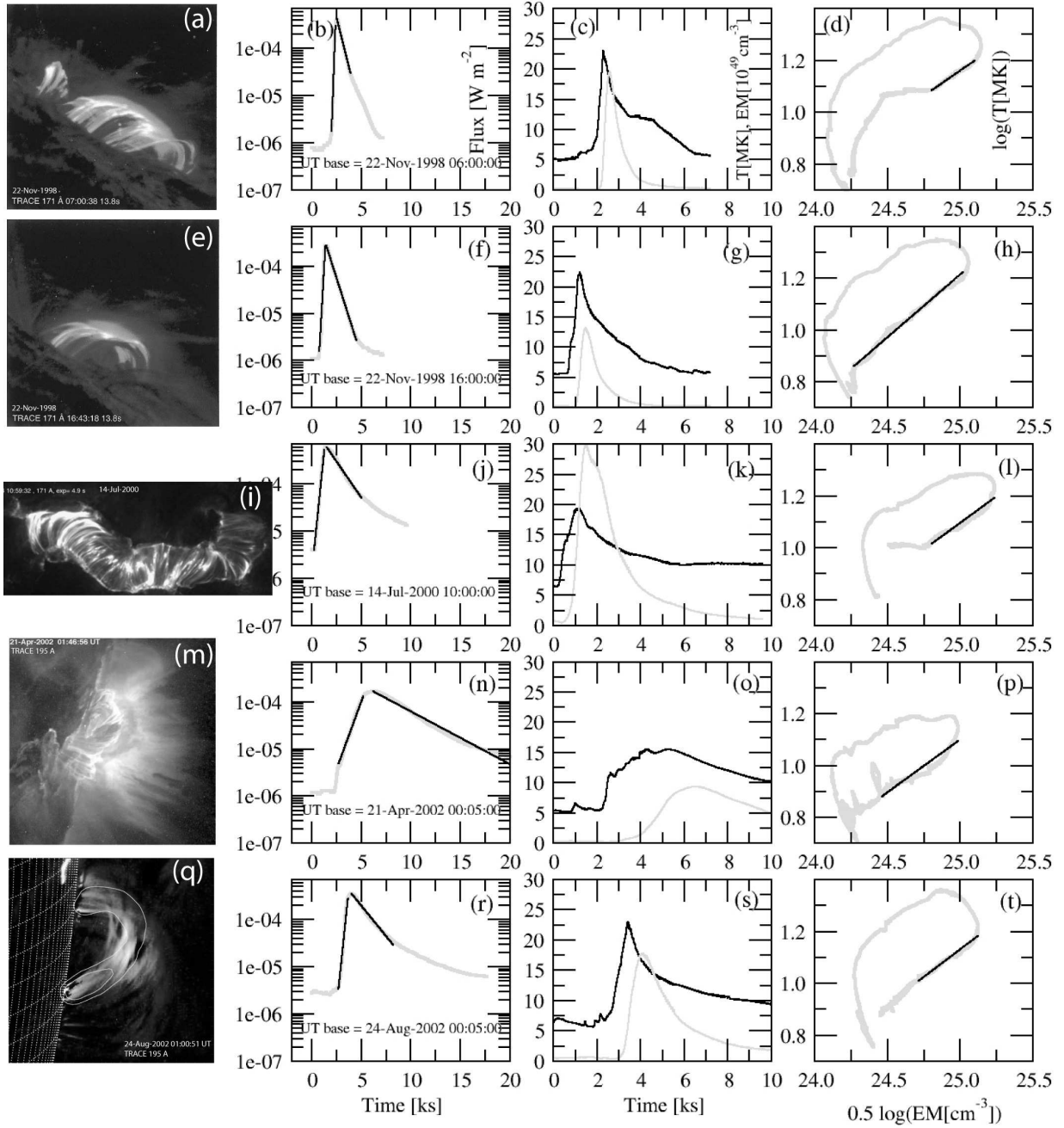}
\caption{Modeling of five solar X-class flares, each represented by a single row of the figure. {\it TRACE} images (171~\AA \ or 195~\AA \ filter) are shown in the first column; fields of view are $100\arcsec \times 100\arcsec$ (panel a), $100\arcsec \times 100\arcsec$ (panel e), $280\arcsec \times 128\arcsec$ (panel i), $210\arcsec \times 210\arcsec$ (panel m), and $110\arcsec \times 110\arcsec$ (panel q). {\it GOES} light-curves ($1-8$~\AA band, gray) with least-squares fits of the rise and decay phases (black) are shown in the second column. Temporal profiles of the plasma temperature (black) and the emission measure (gray) are shown in the third column. Evolution in the $\log T$~--~$\log \sqrt{EM}$ plane (gray) with a fit to the flare decay slope ($\zeta$, black line) is shown in the fourth column. \label{fig_solar_flares}} 
\end{figure}

From the {\it TRACE} solar flare catalog\footnote{\url{http://hea-www.harvard.edu/trace/flare\_catalog/index.html}} we choose only limb flares, where the heights of coronal structures derived from high resolution {\it TRACE} images at EUV are most reliable. Out of the 75 X-class flares listed in the catalog, $\sim 10$ are limb flares, and of those at least the following 4 have been studied in detail: X3.7 and X2.5 flares on November 22, 1998 \citep{Warren00};  X1.5 flare on April 21, 2002 \citep{Gallagher02}; and X3.1 flare on August 24, 2002 \citep{Li05,Reznikova09}. To these 4 flares we add the well studied Bastille Day (14 July 2000) near-disk center flare \citep{Aschwanden01}. All these five flare events are associated with arcades of multiple loops (Figure \ref{fig_solar_flares}). At least for the two events on 21-Apr-02 \citep{Gallagher02} and 24-Aug-2002 \citep{Li05}, ``rising'' flaring loop systems are reported and are believed to be associated with the formation of new loops as magnetic reconnection progresses upward. The heights of the EUV loops inferred from {\it TRACE} images, $H$, are given in Column~9 of Table~\ref{tbl_solar_flares}. We assume that the X-ray emitting loops for the five events have heights comparable to or higher than the observed EUV loops; this assumption is generally expected due to effects of loop cooling and shrinkage, and is found for both 22-Nov-98 flares \citep[Figure~3 in][]{Warren00}. As a sanity check of our {\it GOES} X-ray flare analysis, presented below, the 3 brightest M-class flares from the original testbed sample of R97 are also analysed.

The {\it GOES} X-ray data have been analysed using the {\it GOES} software tools within an IDL based data analysis software package for Solar Physics, {\it SolarSoft} \citep{Freeland98}\footnote{Descriptions and codes for {\it SolarSoft}, and {\it GOES} IDL User Guide can be found at \url{http://www.lmsal.com/solarsoft/} and \url{http://hesperia.gsfc.nasa.gov/$\sim$kim/goes\_software/goes.html}, respectively.}. Background subtraction for these powerful solar flares is ignored. Plasma temperature and emission measure profiles are derived by applying the filter-ratio method \citep[{\it goes\_chianti\_tem.pro} program from][]{White05} to the observed {\it GOES} light curves in the ($0.5-4$~\AA) and ($1-8$~\AA) bands. 

Light curve timescales, observed peak plasma temperature and emission measure, and slope $\zeta$ of the trajectory in the $\log T$~--~$\log \sqrt{EM}$ diagram are shown in Figure~\ref{fig_solar_flares} and are given in Columns~$2-6$ of Table~\ref{tbl_solar_flares}. Estimates on loop half-length from the rise, $L_{rise}$, and decay, $L_{decay}$, phases of light curves, using the single loop approach, are also tabulated (Columns~$7-8$) and compared to the heights of the corresponding EUV loops, $H$ (Column~9). Loop half-length from the rise timescale of the light-curve, assuming an explosive chromospheric evaporation, is given as $L_{rise}\rm{[cm]} = \tau_{rise}\rm{[s]} \times 1.5 \times10^{4} \sqrt{T_{obs,pk}\rm{[K]}}$ (\S \ref{sec_single_vs_multiple}). Loop half-length from the decay phase of the light-curve is calculated using the R97 method (\S \ref{sec_loop_length}) with the instrument-dependent parameters in the calibration formulas for $F(\zeta)$ and $T_{pk}$ taken for {\it GOES9} from Table~A.1 of \citet{Reale07}. Formal statistical errors on derived {\it GOES} temperature and emission measure are expected to be $\la 4$\% \citep{White05}, leading to statistical errors on the slope $\zeta$ of $\la 3$\% and errors on the loop half-length derived from the decay phase of the light curve $L_{decay}$ of $\la 30$\%. 

For the three brightest M-class flares from the original sample of R97, comparison of their flare properties derived using {\it Yohkoh/SXT} data (R97) with those inferred from our {\it GOES} analysis shows that {\it GOES} peak plasma temperatures are systematically higher by $30$\%, slope $\zeta$ differences are within $30$\%, and loop half-length $L_{decay}$ differences are within $50$\% (with a formal statistical precision of $\la 30$\% on individual flares). Effects of neglecting background emission in {\it GOES} data are expected to be more pronounced for M-class rather than for X-class flares.

The result of our analysis is that for the five X-class flares associated with arcades of multiple loops (Figure~\ref{fig_solar_flares}), loop half-lengths $L_{decay}$ (or more precisely loop heights, $0.64 \times L_{decay}$, assuming semi-circular loops stand vertically on the solar surface) derived from disk-integrated {\it GOES} data using the single loop approach of R97 are shorter or comparable to the heights of the EUV loops, $H$, measured from a direct inspection of high-resolution {\it TRACE} images (Table~\ref{tbl_solar_flares}). It is important to note that the application of the single-loop approach to multi-loop flares is a priori incorrect, and the similarity in loop length values can occur by chance (\S \ref{sec_single_vs_multiple}) or because of a special physical phenomenon taking place, namely the presence of a dominant X-ray emitting structure. The latter does not seem to apply to the Bastille Day flare (see below). Meanwhile, the loop half-lengths derived from the rise phases of the {\it GOES} light curves with the assumption of chromospheric evaporation in a single loop, $L_{rise}$, (or their semi-circular heights, $0.64 \times L_{rise}$) are systematically larger by a factor of $4-7$ compared to the measured heights of the EUV loops. It is also important to note two other aspects of our analyses. First, for all five flares the inferred slope of the trajectory in the $\log T$~--~$\log \sqrt{EM}$ diagram is $\zeta \sim 0.4$, indicating that heating is the dominant process in the decay phase of a flare (\S \ref{sec_single_vs_multiple}). Second, the effect of time delay between temperature and emission measure peaks is clearly observed in all 5 multi-loop X-class flares (Figure~\ref{fig_solar_flares}). This confirms the general expectation that the effect (described in \S \ref{sec_trs}) is applicable to a multi-loop flare case as well as to a single loop. In a multi-loop case the effect can be explained by the principle of linear superposition of flares from individual loops to produce a flare for a complete event \citep[e.g.,][]{Aschwanden01}. Below we apply the principle of superposition to the Bastille Day flare.

If individual loops in a flaring loop system are ignited nearly simultaneously and {\it are subject to a similar evolution, which is rather a quite rare condition,} (or a single loop dominates the flare X-ray emission), then for the complete event, according to the superposition principle, the flare temperature profile (including its peak plasma temperature), the flare light curve shape (specifically the e-folding decay timescale), and the slope $\zeta$ on the $\log T$~--~$\log \sqrt{EM}$ diagram should be similar to those of individual loops. In such case, for the complete event the equation of R97 would yield loop length similar to those of individual loops (or a single dominant loop). Is this the case for the 5 solar X-class flares studied here? At least for the best studied case, the Bastille Day flare, we are unable to collect strong evidences in support of such a picture. 

The plasma temperature profile of the Bastille Day event (Figure \ref{fig_solar_flares}) has a major peak (near 10:18:20~UT) and two bumps seen along the rise (near 10:10:00~UT) and the decay (near 10:28:20~UT) phases of the profile, respectively. The classic time delay from temperature to emission measure is seen at the major peak ($\Delta T \sim 340$~seconds) and decay-phase bump ($\Delta T \sim 200$~seconds). According to the TRACE images \citep[Figure 7 in][]{Aschwanden01} and the movie\footnote{The TRACE movie of the Bastille Day flare can be found e.g., at \url{http://www.suntrek.org/gallery/gallery-magnetic-sun.shtml}.}, the rise-phase temperature bump is likely related to the appearance of several EUV loops at the westernmost edge of the arcade (near 10:10:00~UT), the major temperature/emission measure peaks are associated with $\sim 20$ EUV loops \citep{Aschwanden01} that start appearing at about 10:14:00~UT (and last up to 11:00:00~UT), and the later temperature/emission measure bumps are likely related to numerous loops at the eastern part of the arcade that produce weaker flares appearing around 10:38:00~UT.

Using the principle of superposition, we simulate profiles of the complete event as $EM_{tot}(t) = \sum{EM_{i}(t)}$ and $T_{tot}(t) = \sum{(T_{i}(t) \times EM_{i}(t))}/\sum{EM_{i}(t)}$ where the summation is over the 20 EUV loops associated with the major peak. Here we assume that the $T_{i}(t)$ and $EM_{i}(t)$ profiles for individual loop events have shapes similar to the profiles of the observed complete event (Figure \ref{fig_solar_flares}). This simplistic simulation shows that if the loops fire nearly simultaneously (within the time period less than the observed $T-EM$ peak delay of 340~seconds, or even more precisely less than 2 minute period), then the temperature profile (both the shape and the peak), the shape of the emission measure profile, as well as the $T-EM$ peak delay of the complete event look similar to those of individual loop events. Otherwise, for a complete event the simulation predicts broader shapes of the temperature and emission measure profiles and longer $T-EM$ peak delay than those of individual loop events, e.g., the delay becomes twice longer, if the loops fire sequentially (and evenly spaced in time) within the 10 minute time period. Thus, the ignition of 20 loops (that are subject to a similar evolution) within the 2 minute time period might mimic a single dominant loop. Notice, however, that in our analysis we make an assumption that the profiles for individual loop events have shapes similar to the observed profiles of the complete event. Many other solutions must exist. For example, using the principle of superposition \citet{Aschwanden01} show that individual loop events not firing near-simultaneously, with temperature and emission measure profiles much shorter than for the complete event, can produce the observed profiles of the complete event.

One might also argue that EUV images might not provide a true picture of an X-ray flare event. The X-ray band Yohkoh/SXT image of the Bastille Day flare corresponding to the major EUV peak \citep[Figure 3 in][]{Aschwanden01} shows a single bright feature (spanning $20 \times 40$ square arcseconds) within the more extended weaker emission, which may be evidence that a few localized EUV-like loops\footnote{In the EUV data \citep[Figure 8 in][]{Aschwanden01} we estimate that a few to several EUV loops can fit within the apparent footprint of the bright X-ray feature.} dominate the X-ray emission \citep{Reale04}. However, since only a single instant X-ray image \citep[Figure 3 in][]{Aschwanden01} is available, it is unclear how the X-ray emission evolves spatially during the early phase of the flare.

Assuming the presence of a dominant X-ray emitting structure (a single loop or multiple similar loops firing simultaneosly) during the Bastille Day flare we can apply the R97 approach strictly to the initial decay segment of the X-ray light curve \citep{Reale04}. The slope of the initial decay (prior to the bump at the late decay phase) on the $\log T$~--~$\log \sqrt{EM}$ diagram is $\zeta \sim 2$, much steeper than the $\zeta = 0.41$ obtained for the entire decay phase of the light curve, leading to an unrealistically large loop size of 200,000~km, which is a factor of 10 larger than the loop lengths measured directly from the EUV images.  Considering the above, it seems to us that there is no clear evidence in support of a dominant X-ray emitting structure in the Bastille Day flare.

\clearpage


\clearpage

\begin{deluxetable}{ccc}
\centering \rotate \tabletypesize{\tiny} \tablewidth{0pt}
\tablecolumns{3}
\tablecaption{X-ray Models for the Low Temperature Components \label{tbl_ch_models}}
\tablehead{

\multicolumn{1}{c}{Parameters} &
\multicolumn{1}{c}{Model~1} &
\multicolumn{1}{c}{Model~2} \\
                                
\multicolumn{1}{c}{\hrulefill} &  
\multicolumn{1}{c}{\hrulefill} &
\multicolumn{1}{c}{\hrulefill} \\

\colhead{(1)} & \colhead{(2)} &
\colhead{(3)}}

\startdata
$N_H$~($10^{21}$~cm$^{-2}$) & 1.3      & 3.2                \\
$kT_1$~(keV)      & 0.7               & 0.3               \\
$EM_1$~($10^{52}$~cm$^{-3}$) & 4.1      & 19.1              \\
$kT_2$~(keV)      & 1.9               & 0.9               \\
$EM_2$~($10^{52}$~cm$^{-3}$) & 6.6      & 8.3              \\
$L_X$~($10^{30}$~erg~s$^{-1}$) & 1.0    & 1.7             \\
\enddata

\end{deluxetable}

\clearpage

\begin{deluxetable}{ccccccccccc}
\centering \rotate \tabletypesize{\tiny} \tablewidth{0pt}
\tablecolumns{11}
\tablecaption{Time-Resolved Spectroscopy \label{tbl_trs}}
\tablehead{

\multicolumn{3}{c}{Time Segment} &
\multicolumn{4}{c}{Using Model~1} &
\multicolumn{4}{c}{Using Model~2} \\
                                
\multicolumn{3}{c}{\hrulefill} &  
\multicolumn{4}{c}{\hrulefill} &
\multicolumn{4}{c}{\hrulefill} \\

\colhead{Seg} & \colhead{$t_0$} &
\colhead{$\Delta t$} & \colhead{$T_{obs}$} & \colhead{$EM$} & \colhead{$L_{X,0.5\_8}$} &
\colhead{$L_{X,0.01\_50}$} & \colhead{$T_{obs}$} & \colhead{$EM$} & \colhead{$L_{X,0.5\_8}$} &
\colhead{$L_{X,0.01\_50}$} \\

\colhead{} & \colhead{(ks)} &
\colhead{(ks)} & \colhead{(MK)} & \colhead{($10^{53}$~cm$^{-3}$)} & \colhead{($10^{30}$~erg~s$^{-1}$)} &
\colhead{($10^{30}$~erg~s$^{-1}$)} & \colhead{(MK)} & \colhead{($10^{53}$~cm$^{-3}$)} & \colhead{($10^{30}$~erg~s$^{-1}$)} & \colhead{($10^{30}$~erg~s$^{-1}$)}
 \\

\colhead{(1)} & \colhead{(2)} &
\colhead{(3)} & \colhead{(4)} & \colhead{(5)} & \colhead{(6)} &
\colhead{(7)} & \colhead{(8)} & \colhead{(9)} & \colhead{(10)} & \colhead{(11)}}

\startdata
 1 &  6.74 & 11.17 & 58.13$_{-12.28}^{+18.80}$ & 1.05$_{-0.09}^{+0.09}$ & 1.43 & 2.03 & 47.67$_{- 7.91}^{+11.20}$ & 1.44$_{-0.12}^{+0.13}$ & 1.84 & 2.57\\
 2 &  7.84 & 11.14 & 71.31$_{-15.71}^{+25.37}$ & 1.03$_{-0.08}^{+0.08}$ & 1.48 & 2.18 & 54.34$_{- 8.81}^{+12.22}$ & 1.41$_{-0.11}^{+0.11}$ & 1.89 & 2.66\\
 3 &  8.83 & 10.88 & 60.44$_{-12.40}^{+20.93}$ & 1.08$_{-0.09}^{+0.09}$ & 1.50 & 2.13 & 49.16$_{- 7.64}^{+11.56}$ & 1.48$_{-0.12}^{+0.12}$ & 1.91 & 2.67\\
 4 &  9.83 & 10.68 & 67.81$_{-13.70}^{+22.63}$ & 1.10$_{-0.08}^{+0.08}$ & 1.58 & 2.29 & 53.70$_{- 8.22}^{+11.20}$ & 1.50$_{-0.11}^{+0.11}$ & 1.99 & 2.80\\
 5 & 10.83 & 10.55 & 63.67$_{-12.82}^{+19.68}$ & 1.14$_{-0.08}^{+0.09}$ & 1.61 & 2.31 & 49.79$_{- 7.14}^{+10.37}$ & 1.57$_{-0.11}^{+0.12}$ & 2.04 & 2.84\\
 6 & 11.88 & 10.52 & 63.34$_{-12.27}^{+18.66}$ & 1.17$_{-0.08}^{+0.09}$ & 1.64 & 2.35 & 49.81$_{- 6.96}^{+ 9.85}$ & 1.59$_{-0.11}^{+0.12}$ & 2.07 & 2.89\\
 7 & 12.74 & 10.15 & 65.56$_{-13.47}^{+21.45}$ & 1.21$_{-0.09}^{+0.09}$ & 1.72 & 2.48 & 49.71$_{- 7.34}^{+10.43}$ & 1.66$_{-0.12}^{+0.13}$ & 2.16 & 3.02\\
 8 & 13.65 &  9.95 & 66.82$_{-13.43}^{+20.68}$ & 1.29$_{-0.09}^{+0.10}$ & 1.84 & 2.67 & 50.41$_{- 7.49}^{+10.28}$ & 1.75$_{-0.12}^{+0.13}$ & 2.29 & 3.20\\
 9 & 14.56 &  9.77 & 61.44$_{-11.69}^{+16.90}$ & 1.32$_{-0.10}^{+0.10}$ & 1.84 & 2.63 & 46.86$_{- 6.85}^{+ 8.70}$ & 1.82$_{-0.13}^{+0.14}$ & 2.31 & 3.22\\
10 & 15.44 &  9.61 & 61.38$_{-10.88}^{+15.38}$ & 1.39$_{-0.09}^{+0.10}$ & 1.93 & 2.75 & 47.22$_{- 6.38}^{+ 8.01}$ & 1.89$_{-0.13}^{+0.13}$ & 2.41 & 3.36\\
11 & 16.34 &  9.46 & 55.79$_{- 9.60}^{+13.97}$ & 1.41$_{-0.11}^{+0.11}$ & 1.90 & 2.68 & 43.70$_{- 5.84}^{+ 7.47}$ & 1.94$_{-0.15}^{+0.15}$ & 2.39 & 3.33\\
12 & 17.24 &  9.37 & 52.26$_{- 8.52}^{+12.51}$ & 1.46$_{-0.11}^{+0.11}$ & 1.93 & 2.71 & 41.75$_{- 5.06}^{+ 7.12}$ & 2.01$_{-0.15}^{+0.15}$ & 2.43 & 3.39\\
13 & 18.00 &  9.02 & 54.69$_{- 8.38}^{+12.12}$ & 1.61$_{-0.11}^{+0.11}$ & 2.15 & 3.04 & 43.26$_{- 5.30}^{+ 6.52}$ & 2.18$_{-0.15}^{+0.15}$ & 2.67 & 3.73\\
14 & 18.81 &  8.83 & 54.86$_{- 8.32}^{+11.72}$ & 1.60$_{-0.10}^{+0.10}$ & 2.15 & 3.03 & 43.42$_{- 5.31}^{+ 6.43}$ & 2.17$_{-0.14}^{+0.14}$ & 2.66 & 3.72\\
15 & 19.57 &  8.58 & 58.39$_{- 9.76}^{+13.26}$ & 1.66$_{-0.11}^{+0.11}$ & 2.27 & 3.22 & 45.30$_{- 6.00}^{+ 7.18}$ & 2.24$_{-0.15}^{+0.15}$ & 2.79 & 3.90\\
16 & 20.35 &  8.42 & 70.39$_{-12.80}^{+18.33}$ & 1.71$_{-0.10}^{+0.10}$ & 2.47 & 3.61 & 50.82$_{- 5.97}^{+ 9.24}$ & 2.28$_{-0.14}^{+0.14}$ & 2.98 & 4.17\\
17 & 21.09 &  8.23 & 70.29$_{-12.45}^{+17.88}$ & 1.76$_{-0.10}^{+0.10}$ & 2.53 & 3.70 & 50.09$_{- 5.90}^{+ 8.41}$ & 2.35$_{-0.14}^{+0.15}$ & 3.06 & 4.28\\
18 & 21.73 &  7.86 & 68.15$_{-11.76}^{+17.20}$ & 1.90$_{-0.11}^{+0.11}$ & 2.72 & 3.96 & 49.90$_{- 6.43}^{+ 8.68}$ & 2.52$_{-0.15}^{+0.16}$ & 3.27 & 4.58\\
19 & 22.39 &  7.61 & 69.42$_{-12.40}^{+17.93}$ & 1.98$_{-0.12}^{+0.12}$ & 2.85 & 4.16 & 49.68$_{- 6.38}^{+ 8.09}$ & 2.64$_{-0.15}^{+0.17}$ & 3.42 & 4.79\\
20 & 23.07 &  7.45 & 91.91$_{-17.13}^{+25.15}$ & 2.04$_{-0.10}^{+0.10}$ & 3.10 & 4.84 & 60.38$_{- 7.93}^{+10.95}$ & 2.65$_{-0.14}^{+0.14}$ & 3.66 & 5.22\\
21 & 23.74 &  7.27 & 78.15$_{-15.31}^{+21.27}$ & 2.09$_{-0.12}^{+0.12}$ & 3.09 & 4.62 & 53.65$_{- 7.23}^{+ 9.46}$ & 2.76$_{-0.17}^{+0.17}$ & 3.68 & 5.17\\
22 & 24.37 &  7.09 & 80.14$_{-15.16}^{+26.27}$ & 2.16$_{-0.12}^{+0.13}$ & 3.22 & 4.84 & 55.60$_{- 7.99}^{+10.61}$ & 2.84$_{-0.18}^{+0.18}$ & 3.82 & 5.40\\
23 & 25.00 &  6.92 & 76.59$_{-14.66}^{+20.50}$ & 2.23$_{-0.13}^{+0.13}$ & 3.29 & 4.89 & 53.18$_{- 6.78}^{+ 9.40}$ & 2.93$_{-0.18}^{+0.18}$ & 3.89 & 5.48\\
24 & 25.57 &  6.66 & 77.69$_{-14.79}^{+20.14}$ & 2.37$_{-0.13}^{+0.14}$ & 3.50 & 5.22 & 53.71$_{- 7.09}^{+ 9.24}$ & 3.10$_{-0.19}^{+0.19}$ & 4.13 & 5.81\\
25 & 26.20 &  6.60 & 65.84$_{- 9.85}^{+14.42}$ & 2.41$_{-0.13}^{+0.13}$ & 3.42 & 4.93 & 48.65$_{- 5.91}^{+ 7.19}$ & 3.16$_{-0.18}^{+0.19}$ & 4.06 & 5.68\\
26 & 26.85 &  6.58 & 65.24$_{- 9.00}^{+14.72}$ & 2.50$_{-0.14}^{+0.14}$ & 3.54 & 5.11 & 48.03$_{- 6.02}^{+ 7.08}$ & 3.31$_{-0.20}^{+0.21}$ & 4.23 & 5.91\\
27 & 27.52 &  6.60 & 68.63$_{- 9.90}^{+14.03}$ & 2.51$_{-0.13}^{+0.13}$ & 3.60 & 5.23 & 48.92$_{- 5.50}^{+ 6.73}$ & 3.32$_{-0.17}^{+0.19}$ & 4.28 & 5.98\\
28 & 28.13 &  6.50 & 62.48$_{- 9.75}^{+12.28}$ & 2.48$_{-0.14}^{+0.15}$ & 3.47 & 4.97 & 45.56$_{- 5.78}^{+ 6.48}$ & 3.31$_{-0.20}^{+0.20}$ & 4.15 & 5.80\\
29 & 28.81 &  6.57 & 58.08$_{- 8.99}^{+11.20}$ & 2.44$_{-0.15}^{+0.15}$ & 3.33 & 4.73 & 43.58$_{- 5.12}^{+ 6.16}$ & 3.24$_{-0.20}^{+0.20}$ & 3.98 & 5.56\\
30 & 29.50 &  6.64 & 50.12$_{- 5.96}^{+ 9.17}$ & 2.52$_{-0.15}^{+0.15}$ & 3.28 & 4.59 & 39.54$_{- 4.46}^{+ 5.31}$ & 3.35$_{-0.21}^{+0.23}$ & 3.96 & 5.53\\
31 & 30.16 &  6.62 & 43.26$_{- 4.47}^{+ 5.44}$ & 2.64$_{-0.14}^{+0.14}$ & 3.23 & 4.52 & 34.68$_{- 3.17}^{+ 3.45}$ & 3.55$_{-0.19}^{+0.19}$ & 3.95 & 5.58\\
32 & 30.85 &  6.68 & 45.28$_{- 5.88}^{+ 6.77}$ & 2.63$_{-0.16}^{+0.16}$ & 3.29 & 4.59 & 35.74$_{- 3.79}^{+ 4.07}$ & 3.53$_{-0.21}^{+0.21}$ & 3.98 & 5.60\\
33 & 31.62 &  6.87 & 38.89$_{- 4.46}^{+ 5.25}$ & 2.43$_{-0.14}^{+0.15}$ & 2.85 & 3.99 & 31.57$_{- 2.77}^{+ 3.28}$ & 3.32$_{-0.19}^{+0.20}$ & 3.54 & 5.05\\
34 & 32.36 &  6.98 & 39.16$_{- 4.59}^{+ 6.00}$ & 2.47$_{-0.15}^{+0.16}$ & 2.91 & 4.07 & 31.83$_{- 3.10}^{+ 3.87}$ & 3.36$_{-0.21}^{+0.21}$ & 3.60 & 5.12\\
35 & 33.21 &  7.30 & 38.58$_{- 4.75}^{+ 6.68}$ & 2.25$_{-0.15}^{+0.15}$ & 2.63 & 3.68 & 30.95$_{- 2.93}^{+ 4.46}$ & 3.09$_{-0.21}^{+0.19}$ & 3.27 & 4.67\\
36 & 33.99 &  7.40 & 37.32$_{- 4.38}^{+ 5.69}$ & 2.25$_{-0.14}^{+0.14}$ & 2.59 & 3.63 & 30.05$_{- 2.69}^{+ 3.78}$ & 3.10$_{-0.19}^{+0.18}$ & 3.22 & 4.63\\
37 & 34.74 &  7.41 & 36.14$_{- 4.38}^{+ 5.11}$ & 2.21$_{-0.14}^{+0.14}$ & 2.51 & 3.53 & 29.56$_{- 2.71}^{+ 3.37}$ & 3.05$_{-0.18}^{+0.18}$ & 3.15 & 4.54\\
38 & 35.64 &  7.74 & 34.72$_{- 4.15}^{+ 4.51}$ & 2.11$_{-0.13}^{+0.13}$ & 2.35 & 3.32 & 28.59$_{- 2.58}^{+ 2.77}$ & 2.95$_{-0.17}^{+0.17}$ & 3.00 & 4.35\\
39 & 36.58 &  8.06 & 37.85$_{- 4.48}^{+ 6.08}$ & 1.99$_{-0.13}^{+0.12}$ & 2.30 & 3.23 & 30.28$_{- 2.72}^{+ 4.05}$ & 2.78$_{-0.18}^{+0.16}$ & 2.91 & 4.17\\
40 & 37.50 &  8.28 & 36.73$_{- 4.78}^{+ 6.24}$ & 1.87$_{-0.13}^{+0.13}$ & 2.14 & 3.01 & 30.09$_{- 2.94}^{+ 4.42}$ & 2.59$_{-0.18}^{+0.17}$ & 2.70 & 3.87\\
41 & 38.39 &  8.39 & 30.40$_{- 3.08}^{+ 4.54}$ & 1.98$_{-0.12}^{+0.12}$ & 2.07 & 2.97 & 26.56$_{- 2.70}^{+ 2.62}$ & 2.73$_{-0.16}^{+0.16}$ & 2.68 & 3.94\\
42 & 39.24 &  8.44 & 30.40$_{- 3.70}^{+ 6.17}$ & 1.94$_{-0.15}^{+0.14}$ & 2.04 & 2.92 & 27.15$_{- 3.38}^{+ 3.27}$ & 2.64$_{-0.18}^{+0.18}$ & 2.62 & 3.83\\
43 & 40.27 &  8.82 & 32.02$_{- 4.62}^{+ 4.77}$ & 1.86$_{-0.12}^{+0.14}$ & 1.99 & 2.84 & 27.21$_{- 3.00}^{+ 2.93}$ & 2.57$_{-0.16}^{+0.16}$ & 2.55 & 3.73\\
44 & 41.22 &  8.94 & 30.25$_{- 3.57}^{+ 5.95}$ & 1.74$_{-0.13}^{+0.12}$ & 1.82 & 2.61 & 27.24$_{- 3.18}^{+ 3.14}$ & 2.38$_{-0.16}^{+0.16}$ & 2.36 & 3.46\\
45 & 42.26 &  9.24 & 34.12$_{- 5.35}^{+ 5.51}$ & 1.61$_{-0.12}^{+0.12}$ & 1.77 & 2.51 & 29.10$_{- 3.13}^{+ 4.24}$ & 2.22$_{-0.16}^{+0.16}$ & 2.27 & 3.28\\
\enddata

\tablecomments{Columns 1-3: Segment's sequential number, central time, and duration. Columns 4-7: Using Model~1 for the fixed low temperature plasma component emission, the flare properties inferred from the spectral fits are: the flare plasma temperature and its 1$\sigma$ errors, emission measure and its 1$\sigma$ errors, absorption-``corrected'' luminosities in the $(0.5-8)$ and $(0.01-50)$~keV bands. Columns 8-11: Same as in Columns 4-7, but using Model~2 for the low temperature plasma component emission.}

\end{deluxetable}

\clearpage

\begin{deluxetable}{cccccccccc}
\centering \rotate \tabletypesize{\tiny} \tablewidth{0pt}
\tablecolumns{10}
\tablecaption{Inferred Properties of The Testbed Solar Flares \label{tbl_solar_flares}}
\tablehead{

\colhead{Flare} & \colhead{$\tau_{rise}$} & \colhead{$\tau_{decay}$} & \colhead{Slope $\zeta$} &
\colhead{$T_{obs,pk}$} & \colhead{$EM_{pk}$} & \colhead{$L_{rise}$} & \colhead{$L_{decay}$} & \colhead{$H$} &
\colhead{Reference}\\

\colhead{} & \colhead{(s)} & \colhead{(s)} & \colhead{} &
\colhead{(MK)} & \colhead{($10^{49}$~cm$^{-3}$)} & \colhead{(1000 km)} & \colhead{(1000 km)} & \colhead{(1000 km)} &
\colhead{}\\

\colhead{(1)} & \colhead{(2)} &
\colhead{(3)} & \colhead{(4)} & \colhead{(5)} & \colhead{(6)} &
\colhead{(7)} & \colhead{(8)} & \colhead{(9)}& \colhead{(10)}}  

\startdata
22-Nov-98 06:30 X3.7 & 77  & 515 & 0.38 & 23.0 & 19.6& 56 & 1   & $\ga 10$ & 1\\
22-Nov-98 16:10 X2.5 & 119 & 655 & 0.48 & 22.4 & 13.2& 84 & 11  & $\ga 10$ & 1\\
14-Jul-00 10:00 X5.7 & 213 & 1422 & 0.41& 19.3 & 29.8& 140 & 8  & 17.5     & 2\\
21-Apr-02 00:05 X1.5 & 761 & 3889 & 0.41& 15.5 & 9.4 & 453 & 18 & $30-50$  & 3\\
24-Aug-02 00:30 X3.1 & 215 & 1680 & 0.42& 23.0 & 17.8& 154 & 14 & $\ga 15$ & 4, 5\\
02-Nov-91 06:44 M9.1 & 195 & 737  & 0.55& 20.5 & 3.2 & 134 & 18 & \nodata & 6\\
06-Feb-92 03:16 M7.7 & 281 & 955  & 0.93& 19.8 & 2.5 & 189 & 64 & \nodata & 6\\
15-Feb-92 21:29 M5.5 & 277 & 712  & 0.83& 17.5 & 2.1 & 174 & 38 & \nodata & 6\\
\enddata
\tablecomments{Columns 1: Flare date, start time, and {\it GOES} class. Columns 2-3: Flare light-curve rise and decay timescales. Column 4: Inferred slope of the trajectory in the $\log T$~--~$\log \sqrt{EM}$ diagram. Columns 5-6: Flare peak observed plasma temperature and emission measure. Column 7: Loop half-length from the rise timescale of the light-curve, assuming an explosive chromospheric evaporation in a single loop, $L_{rise}\rm{[cm]} = \tau_{rise}\rm{[s]} \times 1.5\times10^{4} \sqrt{T_{obs,pk}\rm{[K]}}$. Column 8: Loop half-length from the decay phase of the light-curve using the single loop approach of R97. Column 9: Apparent loop height from {\it TRACE} images. Column 10: Reference for columns (1) and (9). REFERENCES: (1) \citet{Warren00}, (2) \citet{Aschwanden01}, (3) \citet{Gallagher02}, (4) \citet{Li05}, (5) \citet{Reznikova09}, (6) R97.}
\end{deluxetable}

\clearpage


\clearpage

\clearpage

\clearpage

\clearpage

\clearpage

\clearpage

\clearpage

\clearpage

\clearpage

\clearpage

\clearpage

\clearpage

\end{document}